\documentclass[onefignum,onetabnum]{siamonline220329}

\usepackage{anyfontsize}
\usepackage{amsfonts}
\usepackage{graphicx}
\usepackage{amsmath}
\usepackage{array}
\usepackage{caption}
\usepackage[T1]{fontenc}
\usepackage[utf8]{inputenc} 
\usepackage{tempora} 

\usepackage[sort]{cite}
\graphicspath{{./Figure/}}

\usepackage{mathrsfs,xspace,multicol}

\newcommand{\is}[1]{\in\mathbb{#1}}

\makeatletter
\newcommand*{\rom}[1]{\expandafter\@slowromancap\romannumeral #1@}
\makeatother
\usepackage{url}
\usepackage{ifthen}
\usepackage[scaled=.8]{beramono}
\usepackage{pifont}
\usepackage{euscript}
\usepackage{cases,balance}
\usepackage{dsfont}
\usepackage[inline, shortlabels]{enumitem}
\setlist{itemjoin ={,\enspace}, itemjoin* = {, and\enspace}}

\usepackage{algorithmic}
\usepackage{algorithm}
\usepackage{wrapfig}

\usepackage[percent]{overpic}

\newcommand{\bpara}[1]		{\smallskip \noindent {\bf #1}}

\definecolor{modulo}{RGB}{72,0,255}
\definecolor{tof}{RGB}{255,0,63}

\usepackage{hyperref}
\hypersetup{
colorlinks=true,
linkcolor=black,
filecolor=black,      
urlcolor=black,
citecolor = black,
}

\newtheorem*{eg}{Example}
\renewcommand{\paragraph}[1]{\smallskip\noindent{\textit{\textbf{#1}}---}}

\usepackage[normalem]{ulem}

\usepackage[inline, shortlabels]{enumitem}

\setlist{itemjoin ={,\enspace},itemjoin* = {, and\enspace}}

\def\DE					{\stackrel{\rm{def}}{=}}
\def\eg					{\emph{e.g.\xspace}}
\def\ie					{\emph{i.e.\xspace}}

\def\SR                 {{\color{black}SRes}\xspace}
\newcommand{\PMs}		{\mathsf{P}_{\mathbf{r}}}
\newcommand{\QMs}		{\mathsf{Q}_{\mathbf{r}}}

\newcommand{\QMj}		{\mathsf{Q}^{\sqb{j}}_{\mathbf{r}}}
\newcommand{\PMjj}		{\mathsf{P}^{\sqb{j+1}}_{\mathbf{r}}}
\newcommand{\QMjj}		{\mathsf{Q}^{\sqb{j+1}}_{\mathbf{r}}}
\renewcommand{\H}		{\mathsf{H}_{\mathbf{r}}}
\newcommand{\Aj}		{\mathbf{A}^{\sqb{j}}_{\mathbf{r}}}
\newcommand{\Bj}		{\mathbf{B}^{\sqb{j}}_{\mathbf{r}}}
\newcommand{\Rj}		{\mathbf{R}^{\sqb{j}}_{\mathbf{r}}}

\newcommand{\Cj}		{\mathbf{C}^{\sqb{j}}_{\mathbf{r}}}

\newcommand{\xjj}		{\mathbf{X}^{\sqb{j+1}}_{\mathbf{r}}}
\newcommand{\xo}		{\mathbf{X}^{\sqb{0}}_{\mathbf{r}}}
\newcommand{\transp}{{\top}} 
\newcommand{\Hermit}{^{\mkern-1.5mu\mathsf{H}}} 
\newcommand{\zm}[2]		{\xi_{#1}^{#2}}
\def\PO					{\boxed{\mathsf{P1}}}
\def\PT                			{\boxed{\mathsf{P2}}}
\def \dk               {\mathscr{D}_K}
\newcommand{\gram}[1]               {\EuScript{G} \rob{#1}}

\usepackage{tikz}
\usepackage{mathdots}
\usepackage{yhmath}
\usepackage{xcolor}
\usepackage{colortbl}
\usepackage{cancel}
\usepackage{siunitx}
\usepackage{multirow}
\usepackage{gensymb,bm}
\usepackage{tabularx}
\usepackage{extarrows}
\usepackage{booktabs}
\usetikzlibrary{fadings}
\usetikzlibrary{patterns}
\usetikzlibrary{shadows.blur}
\usetikzlibrary{shapes}

\def\N					{\mathbb N}
\def\Z					{\mathbb Z}
\def\R					{\mathbb R}
\def\C					{\mathbb C}

\def\iZ					{\in \mathbb Z}
\def\iR					{\in \mathbb R}
\def\iC					{\in \mathbb C}
\def\td{$2$D }
\def\ttd{$3$D }
\def\etal{\emph{et al.~}}

\def\e					{{e}}
\def\DE					{\stackrel{\rm{def}}{=}}

\def\BSD                    {BSD\xspace}

\def\U					      {\mathbf{U}}
\def\I					      {\mathbf{I}}
\newcommand\Toep[1]				      {\mathbf{T}_{#1}}

\def\l						{\left(}
\def\r						{\right)}

\newcommand\rob[1]			{\l #1 \r}
\newcommand\fig[1]			    {Fig.~\ref{#1}}
\newcommand\secref[1]			{Section \ref{#1}}
\newcommand\tabref[1]			{Table \ref{#1}}
\newcommand\algref[1]			{Algorithm \ref{#1}}

\newcommand\lemmaref[1]			{Lemma \ref{#1}}

\newcommand{\sqb}[1]		{\left[ #1 \right]}

\newcommand{\ft}[1]			{\left[\kern-0.15em\left[#1\right]\kern-0.15em\right]}
\newcommand{\fe}[1]		{\left[\kern-0.30em\left[#1\right]\kern-0.30em\right]}

\newcommand{\df}[3]		{{\partial^{(#1)}_{#3}} #2}

\newcommand{\mat}[1]		{\mathbf{#1}}

\newcommand{\mse}[2]		{\EuScript{E}{\rob{\mat{#1},\mat{#2}}}}
\newcommand{\msem}[2]                   		{\EuScript{E}{\rob{\mat{#1}_{\mathbf{r}},{\widetilde{\mat{#2}}}_{\mathbf{r}}}}}
\newcommand{\PSNR}[2]                   		{\mathsf{PSNR}{\rob{\mat{#1},{\widetilde{\mat{#2}}_{\mathbf{r}}}}}}
\newcommand{\msep}[2]                   		{\EuScript{E}{\rob{\mat{#1},{\widetilde{\mat{#2}}_{\mathbf{r}}}}}}

\newcommand{\vmat}[2]		{\mathbf{V}_{#1}^{#2}}
\newcommand{\wmat}[2]		{\mathbf{W}_{#1}^{#2}}

\newcommand{\EQc}[1]		{\stackrel{\eqref{#1}}{=}}

\def\pr					{p_{\mathbf{r}}}
\def\sr 				{s_{\mathbf{r}}}
\def\gr 				{g_{\mathbf{r}}}
\def\yr 				{y_{\mathbf{r}}}
\def\qr 				{d_{\mathbf{r}}}
\def\ur 				{u_{\mathbf{r}}}
\def\mur             {\mathbf{u}_{\mathbf{r}}}
\def\bgr             {\breve{g}_{\mathbf{r}}}
\def\bp             {\breve{\varphi}_{\mathbf{r}}}
\def\bsr             {\breve{s}_{\mathbf{r}}}
\def\hgr             {\widehat{g}_{\mathbf{r}}}
\def\hp             {\widehat{\varphi}_{\mathbf{r}}}
\def\mbp		{\breve{\mat{\bm\varphi}}_{\mathbf{r}}}

\def\circov		{\circledast}
\def\fr              {\varphi_{\mathbf{r}}}
\def\bpr             {{\overline \varphi}_{\mathbf{r}}}
\newcommand\fl[1]               {f_{#1}}

\def\mgr 			{\mathbf{g}_{\mathbf{r}}}

\def\mqr 			{\mathbf{d}_{\mathbf{r}}}
\def\mphi 			{{\bm\varphi}_{\mathbf{r}}}
\def\rr 				{r_{\mathbf{r}}}
\def\phir            {\psi_{\mathbf{r}}}
\def\gammar            {\Gamma_{\mathbf{r}}}
\def\taur            {\tau_{\mathbf{r}}}
\def\srf               {SRF\xspace}
\def\p                  {\mathbf{p}_{\mathbf{r}}}
\def\q                  {\mathbf{q}_{\mathbf{r}}}

\def\cosi					{\textsc{CoSI}\xspace}

\newcommand{\id}[1]		{\mathbb{I}_{#1} }
\newcommand{\Lp}[1]{{\mathbf{L}}_{{#1}}}
\newcommand{\lp}[1]{{\ell}_{#1}}
\newcommand{\normt}[3]{ {\| {#1} \|}_{ {\Lp{#2}} \rob{#3}}}

\newcommand{\norm}[1]{\left\| {#1} \right\|}
\newcommand{\abs}[1]{\left| #1\right|}
\newcommand{\inner}[2] {\left\langle {#1,#2} \right\rangle}

\newcommand{\Lap}[2]		{\mathcal{L}_{#1} \rob{#2} }
\newcommand{\eqr}[1]		{\stackrel{\eqref{#1}}{=}}

\newcommand{\jmax}  {j_{\rm max}}

\renewcommand\bar\underline
\renewcommand\hat\widehat
\renewcommand\geq\geqslant
\renewcommand\leq\leqslant
\renewcommand\Psi\ddvmat
\usepackage{stackengine}
\stackMath
\newlength\matfield
\newlength\tmplength
\def\matscale{1.}
\newcommand\dimbox[3]{%
\setlength\matfield{\matscale\baselineskip}%
\setbox0=\hbox{\vphantom{X}\smash{#3}}%
\setlength{\tmplength}{#1\matfield-\ht0-\dp0}%
\fboxrule=1pt\fboxsep=-\fboxrule\relax%
\fbox{\makebox[#2\matfield]{\addstackgap[.5\tmplength]{\box0}}}%
}

\newcommand\matbox[5]{
\stackunder{\dimbox{#1}{#2}{${#5}$}}{\scriptstyle(#3\times #4)}%
}

\renewcommand\tilde\widetilde

\makeatletter
\def\moverlay{\mathpalette\mov@rlay}
\def\mov@rlay#1#2{\leavevmode\vtop{%
\baselineskip\z@skip \lineskiplimit-\maxdimen
\ialign{\hfil$\m@th#1##$\hfil\cr#2\crcr}}}
\newcommand{\charfusion}[3][\mathord]{
#1{\ifx#1\mathop\vphantom{#2}\fi
\mathpalette\mov@rlay{#2\cr#3}
}
\ifx#1\mathop\expandafter\displaylimits\fi}
\makeatother

\usepackage{bbm,amssymb,soul}

\newsiamremark{hypothesis}{Hypothesis}
\crefname{hypothesis}{Hypothesis}{Hypotheses}
\newsiamthm{claim}{Claim}

\headers{SIAM Journal on Imaging Sciences}{Ruiming Guo and Ayush Bhandari}

\title{
Blind Time-of-Flight Imaging:\\ Sparse Deconvolution on the Continuum with Unknown Kernels%
\thanks{Article in press.
\funding{The work of the authors is supported by the UK Research and Innovation council's FLF Program ``Sensing Beyond Barriers via Non-Linearities'' (MRC Fellowship award no.~MR/Y003926/1).}}
}

\author{Ruiming Guo\thanks{The authors are with the Dept. of Electrical and Electronic Engineering, Imperial College London, SW72AZ, UK 
(\email{{\{ruiming.guo,a.bhandari\}@imperial.ac.uk} or ayush@alum.mit.edu}).}
\and Ayush Bhandari\footnotemark[2]
}

\usepackage{amsopn}

\newcommand\rg[1]			{{\color{black}#1}}
\newcommand\rgn[1]			{{\color{black}#1}}
\newcommand\rgnn[1]			{{\color{black}#1}}

\ifpdf
\hypersetup{
pdftitle={SIAM Journal on Imaging Sciences (in press)},
pdfauthor={Ruiming Guo and Ayush Bhandari}
}
\fi

\begin{document}

\maketitle

\begin{abstract} In recent years, computational Time-of-Flight (ToF) imaging has emerged as an exciting and a novel imaging modality that offers new and powerful interpretations of natural scenes, with applications extending to 3D, light-in-flight, and non-line-of-sight imaging. Mathematically, ToF imaging relies on algorithmic super-resolution, as the back-scattered sparse light echoes lie on a finer time resolution than what digital devices can capture. Traditional methods necessitate knowledge of the emitted light pulses or kernels and employ sparse deconvolution to recover scenes. Unlike previous approaches, this paper introduces a novel, blind ToF imaging technique that does not require kernel calibration and recovers sparse spikes on a continuum, rather than a discrete grid. By studying the shared characteristics of various ToF modalities, we capitalize on the fact that most physical pulses approximately satisfy the Strang-Fix conditions from approximation theory. This leads to a new mathematical formulation for sparse super-resolution. Our recovery approach uses an optimization method that is pivoted on an alternating minimization strategy. 
We benchmark our blind ToF method against traditional kernel calibration methods, which serve as the baseline. Extensive hardware experiments across different ToF modalities demonstrate the algorithmic advantages, flexibility and empirical robustness of our approach. We show that our work facilitates super-resolution in scenarios where distinguishing between closely spaced objects is challenging, while maintaining performance comparable to known kernel situations. Examples of light-in-flight imaging and light-sweep videos highlight the practical benefits of our blind super-resolution method in enhancing the understanding of natural scenes.
\end{abstract}

\begin{keywords}
Computational imaging, inverse problems, blind deconvolution, super-resolution and time-of-flight imaging.
\end{keywords}

\begin{MSCcodes}
62H35, 
68U99,
78A46,
94A12.
\end{MSCcodes}

\section{Introduction to Time-of-Flight Imaging}
\label{sec:intro}
The emerging theme of \emph{Computational Sensing and Imaging} or \cosi \cite{Bhandari:2022:B,Bouman:2022:B} has catalyzed never-seen-before capabilities in the context of imaging and vision. Some noteworthy examples
\emph{single-pixel} imaging \cite{Sen:2005:C,Duarte:2008:J},
\emph{non-line-of-sight} imaging \cite{Velten:2012:J,Wu:2021:J}, 
\emph{ultrafast} imaging at a trillion \cite{Velten:2012:J} and a billion \cite{Gao:2014:J} frames per second, and
imaging of \emph{black hole} \cite{EHTC:2019:J}. In all such examples and beyond, the fundamental difference from the traditional viewpoint is the new mindset that altering the forward model of an imaging system enables new capabilities---one can \emph{co-design} hardware and algorithms in pursuit of new advantages that can not be harnessed by optimizing hardware or algorithms alone.

\begin{figure*}[!t]
\centering
\includegraphics[width=\columnwidth]{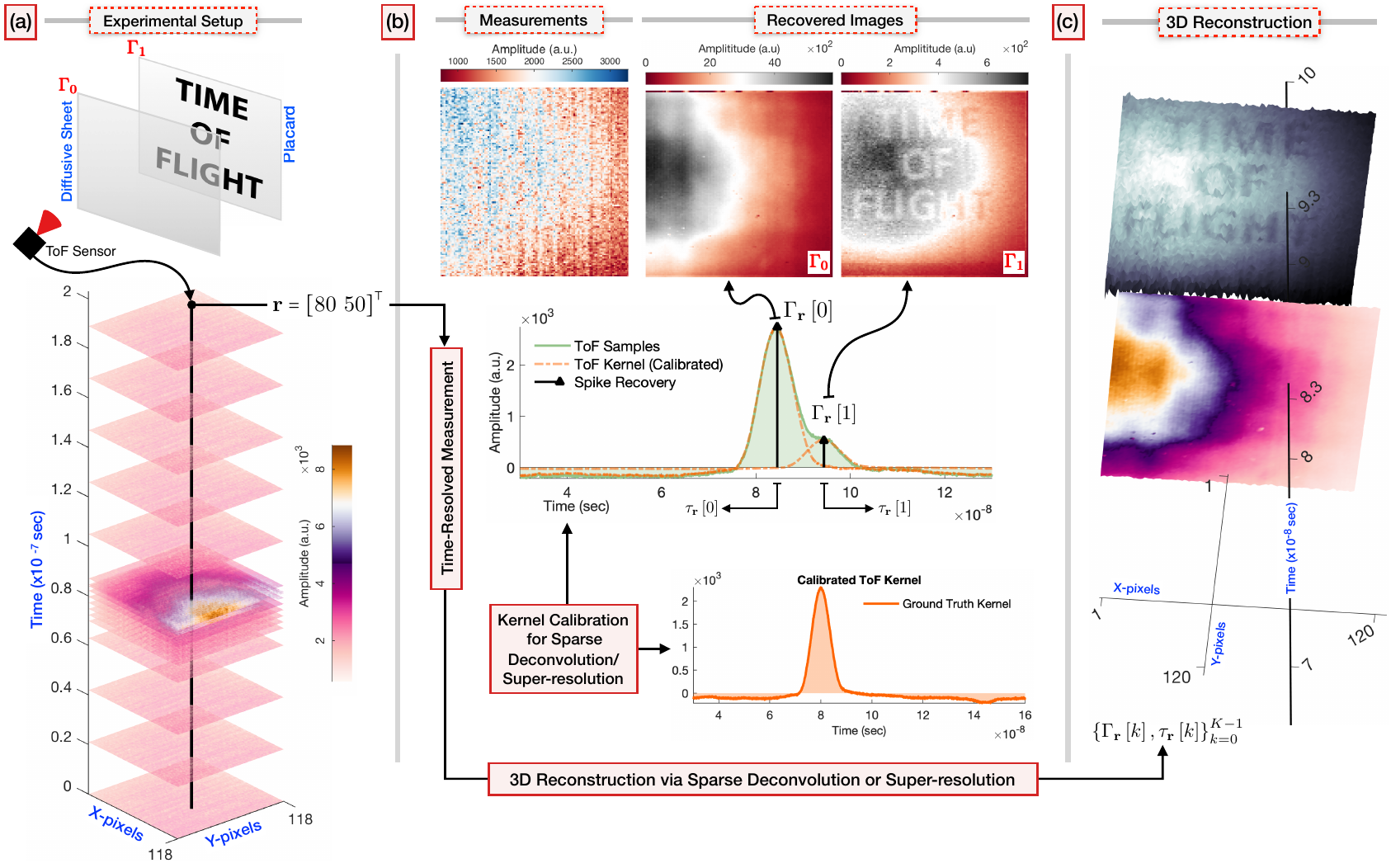}	
\caption{Diffusive Imaging. (a) Experimental setup \cite{Kadambi:2013:J,Bhandari:2016:J}. A ``TIME OF FLIGHT'' placard is hidden by a diffusive semi-translucent sheet between the placard and the ToF sensor. (b) Depth estimation from pixel measurements via sparse deconvolution techniques \cite{Candes:2013:J,Bhandari:2016:J,Bhandari:2016:C}. The ToF kernel (\ie $\fr$ in \eqref{eq:ToF measurements}) is calibrated and known prior to experiments. (c) \ttd scene reconstruction from estimated depth parameters in (b). The main goal of this paper is to recover \ttd scene without pre-known kernel information (refer to \fig{fig:ToFkernel} and \fig{fig:headSR}), aiming to eliminate the necessity for experimental calibration.}
\label{fig:3DToF}
\end{figure*}

In the context of \cosi, the hardware-software co-design approach has been particularly fruitful for time-of-flight (ToF) imaging. A notable characteristic of ToF imaging is that each pixel of the imaging sensor captures a scene-dependent time profile at a specific time resolution, usually on a scale ranging from nanoseconds (ns) to picoseconds (ps). 
This technique, therefore, is also referred to as Time-Resolved Imaging, highlighting its capability to capture detailed temporal information within a scene (see Chapter 5, \cite{Bhandari:2022:B}), as illustrated in \fig{fig:3DToF}.
In contrast to conventional imaging paradigms, ToF imaging offers numerous advantages that were previously unimaginable. This is due to the fact that each ToF measurement comprises two types of images: the conventional \td photograph or amplitude image, and the unconventional time-resolved, depth image.
The advent of ToF imaging technology paves the way for novel methods and applications in various domains, including computer vision \cite{Jarabo:2017:B,Toole:2017:C,Conde:2020:J}, graphics \cite{Wu:2012:C,Kadambi:2013:J,Heide:2013:J}, autonomous vehicles, and bio-medical imaging \cite{Bhandari:2015:J,Satat:2015:J}. Two concrete examples that also serve as an experimental validation of our work (in \secref{sec:exp}) are as follows.
\begin{enumerate}[label = \arabic*)]
\item Diffuse Imaging \cite{Kadambi:2013:J,Heide:2013:J,Bhandari:2016:J,Bhandari:2020:J}. As illustrated in \fig{fig:3DToF}, here ToF imaging enables recovery of objects hidden or obfuscated via diffusive materials, also see \secref{subsubsec:Diffuse Imaging}. 
\item Light-in-Flight (LIF) Imaging \cite{Abramson:1978:J,Heide:2013:J,Velten:2013:J,Kadambi:2013:J,Gariepy:2015:J}. In 1978 \cite{Abramson:1978:J}, N.~Abramson pioneered LIF using a holographic method to record the wavefront of a light pulse as it traveled and scattered off a white screen. In a similar spirit, ToF measurements also enable LIF imaging that involves capturing and reconstructing the trajectory of light as it traverses and interacts with various objects in the scene, leading to completely new pathways for scene understanding. This is another ToF imaging capability, covered in this paper in \secref{subsec:Light-in-Flight Imaging} 
({\href{https://youtu.be/ffkc_z8ogE8}{\texttt{https://youtu.be/ffkc\_z8ogE8}}}).
\end{enumerate}
Depending on the technology and temporal resolution, ToF imaging setups can be broadly classified into three categories:
\begin{enumerate*}[label = \roman*)]
\item Lock-in sensors \cite{Foix:2011:J} operating at nanosecond resolution
\item Single Photon Avalanche Diode (SPAD) detectors \cite{Kirmani:2014:J,Gariepy:2015:J} operating at picosecond resolution
\item Streak tubes \cite{Velten:2012:J} operating at femtosecond resolution.
\end{enumerate*}
Among them, lock-in sensors are notably popular for ToF imaging, largely because they are available as consumer-grade technology, exemplified by \textit{photonic mixer devices} (PMD) and the Microsoft Xbox One's Kinect, making them widely used and accessible.

\begin{figure*}[!t]
\centering
\includegraphics[width=\columnwidth]{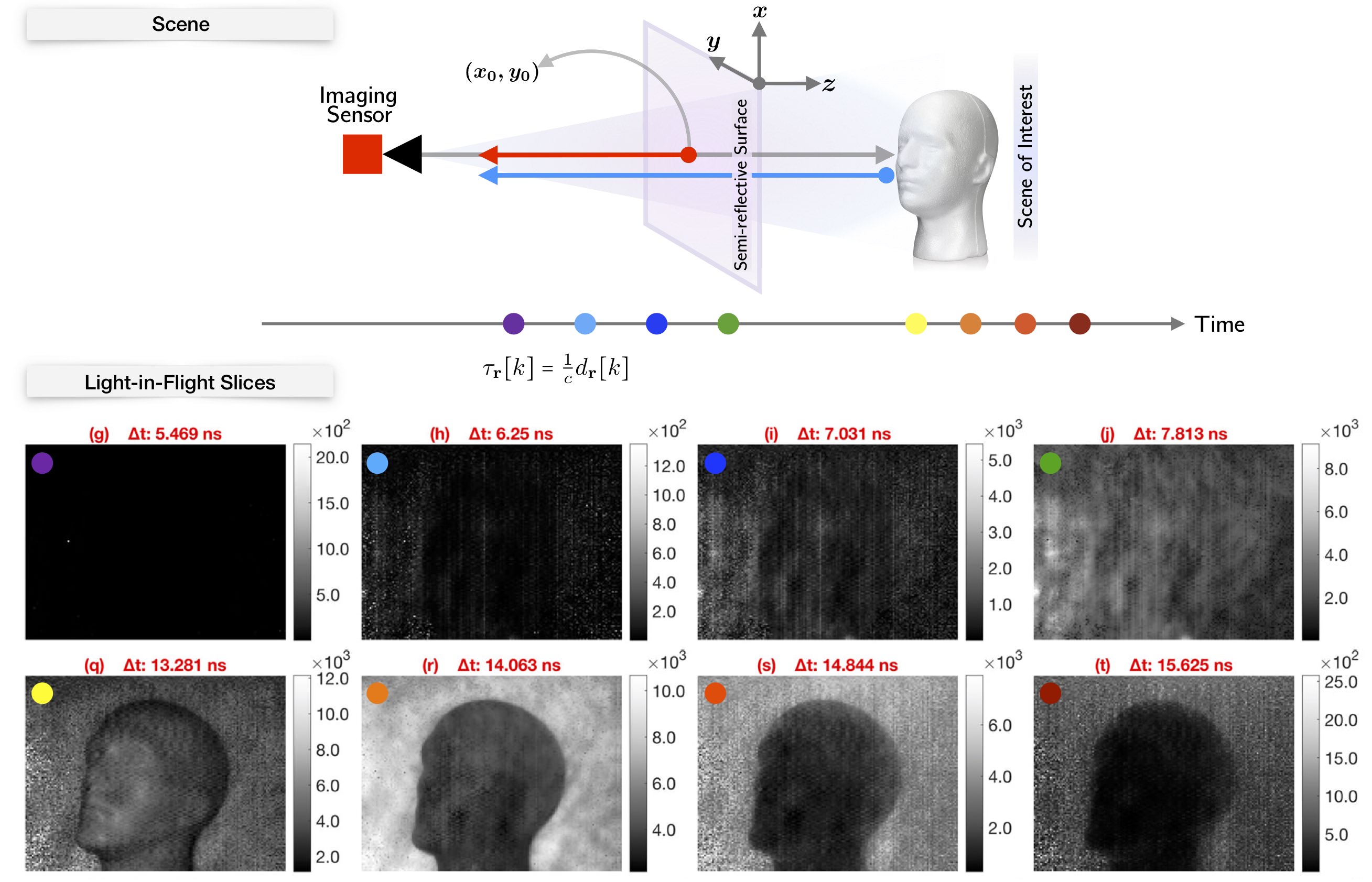}	
\caption{Pipeline for LIF Imaging (see 
\href{https://youtu.be/ffkc_z8ogE8}{\texttt{https://youtu.be/ffkc\_z8ogE8}}). 
Top Row: experimental setup capturing light's interaction with a scene, consisting of a mannequin head positioned between a diffusive surface and a backdrop wall. 
Bottom Row: Using a ToF camera, we visualize the light's progression across different time instances marked by ``$\bullet$'', highlighted by visible light's spectrum colors coding the ascending time sequence. 
Initially, light encounters the diffusive sheet in the early time-slots, \ie (i) (blue) and (j) (green). Light then moves over the mannequin, and ultimately reaches the back wall, as shown in frames (q) (yellow) to (t) (red). The LIF accurately maps the \ttd geometry of the scene. The full LIF time slices are shown in \fig{fig:headLIF}. 
}
\label{fig:LIF}
\end{figure*}

\subsection{Fundamental Role of Sparsity and Super-Resolution in ToF Imaging}
\label{subsec: Sparse ToF Imaging and Related Work}

In conventional 2D photography the scene is rich in its natural representation and described by low-level image features, such as edges, corners and ridges \cite{Elad:2010:B}. The resulting images are sparse in transform domain \eg~DCT or wavelet basis. However, when the same scene is interpreted from the perspective time-scales \cite{Wu:2012:C,Kadambi:2013:J,Heide:2013:J}, the features are fundamentally different and take the form of global and direct delays, inter-reflections and sub-surface scattering. Mathematically, this translates to the fact that scenes are naturally sparse along the temporal dimension. This is also clearly seen from \fig{fig:3DToF} where the time profile along a particular pixel comprises of $2$ Dirac impulses. Clearly, in terms of the Shannon-Nyquist method of digitization, acquiring such scenes would entail exorbitant sampling rates which may be technologically unviable, not to mention how expensive and challenging it may be to implement such systems. That said, realizing that ToF measurements encode sparse features, super-resolution (\SR) or the problem of recovering spikes from filtered kernels \cite{Donoho:1992:J,Candes:2013:J,Poon:2019:J,Catala:2019:J,Denoyelle:2019:J} turns out to be the most appropriate setup for ToF imaging methods \cite{Bhandari:2022:J}. The main advantage of resorting to such a flavor is that non-bandlimited objects such as Dirac impulses can be recovered from filtered measurements, without requiring a sampling criterion \cite{Bhandari:2016:J}. Hence, resorting to \SR formulation proves to be highly beneficial in lowering the hardware constraints.

\subsection{Related Work}

In the existing literature, the ToF imaging methods via \SR formulation can be broadly classified as follows.
{\bf Stochastic Approaches.} The authors in \cite{HernandezMarin:2007:J} proposed a Bayesian framework to find out the \ttd scene parameters via evaluating posterior probability distribution. Adam \etal \cite{Adam:2017:J} utilize Bayesian inference for the recovery of \ttd scene geometry, such as shape, illumination, and albedo. 
{\bf Optimization Approaches.} Kadambi \etal in \cite{Kadambi:2013:J} formulated the time profile recovery with multi-path interference as sparse deconvolution and utilized OMP (orthogonal matching pursuit) algorithm to solve for spike estimation. In \cite{Heide:2013:J}, Heide \etal posed the scene recovery as a temporal-spatial regularization problem, encompassing both spatial and temporal regularized penalties. An alternating minimization strategy was utilized to split the principle problem into two tractable sub-problems via proximal operators. 
{\bf Parametric Approaches.} Built on the sparse representation of ToF measurements, the authors in \cite{Bhandari:2014:Cb,Bhandari:2016:C,Bhandari:2016:J} first modeled the inverse problem of scene recovery as deconvolution, which essentially boils down to spectral estimation problem \cite{DeFigueiredo:1982:J}. Subsequently, Prony's method is employed to super-resolve the parameters of the \ttd scene.

\subsection{Motivation for Blind ToF Imaging}
\label{subsec:motivation}
By comparing and contrasting the mathematical \SR \cite{Donoho:1992:J,Candes:2013:J,Poon:2019:J,Catala:2019:J,Denoyelle:2019:J} and ToF \SR \cite{HernandezMarin:2007:J,Adam:2017:J,Kadambi:2013:J,Heide:2013:J,Bhandari:2016:C,Bhandari:2016:J}  communities, the key takeaways from the existing art can be distilled as follows.
\begin{enumerate}[leftmargin = *]

\item Significance of Kernel Calibration: A vital assumption in any \SR technique \cite{Donoho:1992:J,Candes:2013:J,Poon:2019:J,Catala:2019:J,Denoyelle:2019:J} is the knowledge of the kernel. In the ToF context \cite{HernandezMarin:2007:J,Adam:2017:J,Kadambi:2013:J,Heide:2013:J,Bhandari:2016:C,Bhandari:2016:J}, this translates to kernel calibration prior to conducting experiments (see \fig{fig:3DToF}(b)). Since ToF systems employ active sensing—where the illumination source is separate from the imaging sensor—any alterations in illumination necessitate a re-calibration of the kernel related to illumination. Over the illumination system's lifespan, physical factors like temperature and optical setup changes may cause variability in the kernel. From the perspective of solving an inverse problem, any discrepancy between the calibrated kernel and ToF measurements can significantly degrade the quality of spike estimation, thus limiting the adaptability and reliability of ToF imaging methods. This highlights the importance of developing blind ToF methods that operate without assuming the knowledge of the kernel \cite{Bell:1995:J}.

\item Community Divide: With the exception of parametric approaches \cite{Bhandari:2014:Cb,Bhandari:2016:C,Bhandari:2016:J}, most ToF \SR methods do not yet leverage ``off-the-grid'' model for spikes or Dirac impulses. This means the unknown sparse signal is presumed to ``live'' on a discrete grid, an assumption that may not reflect reality. Conversely, a notable advancement within the mathematical \SR \cite{Candes:2013:J,Poon:2019:J,Catala:2019:J,Denoyelle:2019:J} community has been the ability to recover spikes on the continuum. We believe, this gap in understanding is largely due to the minimal interaction between the two fields. Moreover, state-of-the-art recovery techniques might not directly transfer to ToF imaging, as they are not tailored for handling large-scale measurements, which can extend to tensors. This calls for the creation of efficient algorithms capable of modeling Dirac impulses continuously and managing the voluminous data produced by ToF sensors. Crucially, the effectiveness of these approaches hinges on their validation and bench-marking through real-world experiments.
\end{enumerate}

\subsection{Contributions}
The main goal of this paper is to develop a flexible, robust and efficient blind ToF imaging method that allows for super-resolved \ttd scene recovery in realistic experimental setups, independent of any kernel calibration. We outline the contributions of our work as follows: 
\begin{enumerate}[leftmargin = *,itemsep=2pt, label = $\bullet$]
\item \textbf{Mathematical Model}: Leveraging our experience with ToF imaging \cite{Bhandari:2016:J,Bhandari:2022:B}, we propose a generic model for ToF imaging pipeline in which the Dirac impulses\footnote{We consider continuous-time spikes in terms of distribution.} are modeled on the continuum.

Different from previous works in computational imaging and computer vision/graphics, we assume that the ToF kernel is unknown and is equipped with Strang-Fix properties. This is to say, the shifts of the ToF kernel reproduce exponential-polynomial functions (see \eqref{eq:reproducing kernel}) \cite{Strang:2011:B}. This conceptualization is a key enabler for blind ToF imaging in that Strang-Fix conditions lead to
\begin{enumerate}[leftmargin = *, label = {\roman*})]
\item a flexible choice of unknown kernel, covering the well-known modalities \eg lock-in sensors \cite{Foix:2011:J} and Time-Correlated Single Photon Counting (TCSPC) systems \cite{Pellegrini:2000:J}. 

\item a key continuous-time recovery of sparse spikes.
\end{enumerate}

\item \textbf{Recovery Algorithm}: We design a robust, efficient and scalable algorithm that achieves fill \ttd-stack processing of ToF measurements. We resort to a non-convex optimization scheme to estimate the unknown kernel and \ttd scene parameters via an alternating minimization strategy, which we validate in various experimental setups and datasets.

\item \textbf{Experimental Validation and Benchmarking}: Through a series of $10$ experiments, we benchmark our proposed approach against existing methods with kernel calibration, which serves as a ground truth. Thus showcasing the comparable performance in various real-world settings, such as, 
\begin{enumerate}[leftmargin = *, label = \roman*)]
\item looking through diffusers (see \ref{subsubsec:Diffuse Imaging}) 
\item Large-scale data processing ($120\times 120\times 3968\times 4$ image tensor, see \secref{subsubsec:K=2}).
\item Inter-target separation of $0.20$ cm (see \secref{subsubsec:K=2}).
\item High-order imaging (see \ref{subsubsec:K=3}) and 5) LIF imaging at nanosecond scale (see \secref{subsec:Light-in-Flight Imaging}).
\end{enumerate}
\end{enumerate}
These experiments corroborate the adaptability, robustness and \SR capability of our method.

In the absence of kernel calibration, our setting takes the form of Blind Sparse Deconvolution (\BSD) problem. In the prior art, \BSD has been studied in various flavors including \BSD via multi-channel \cite{Kazemi:2014:J,Wang:2016:J} and single-channel \cite{Kuo:2020:J,Wang:2022:J} methods. 
However, such \BSD approaches do not translate to the ToF context due to 
\begin{enumerate*}[leftmargin = *, label = \uline{\arabic*})]
\item kernel priors \eg incoherence \cite{Li:2019:J}, non-negativity \cite{Perrone:2016:J} and short-support \cite{Wang:2022:J}
\item assumption that Dirac impulses or spikes lie on a grid \cite{Kazemi:2014:J,Repetti:2015:J,Wang:2022:J}
\item the large scale data that arises in ToF imaging \cite{Kadambi:2013:J,Bhandari:2016:J}.
\end{enumerate*}
Finally, we find that prevailing works on \BSD are not designed to handle the ToF pipeline. The lack of experimental validation of such approaches not only creates a gap between theory and practice but also makes bench-marking of these approaches very challenging.

\begin{figure*}[!tb]
\centering
\resizebox{\columnwidth}{!}{
$
\underbrace{\pr}_{\text{Emitted Signal}} \to \underbrace{\boxed{\boxed {\sr}}}_{\text{\srf}} \to \underbrace{\color{black!100} r_{\mathbf{r}} (t) = \int \pr \rob{t'} \sr \rob{t,t'}  dt'}_{\text{Reflected Signal}} \to \underbrace{\boxed{\boxed\psi }}_{\text{IRF}} \to \underbrace{ \color{black!100} \gr\rob{t} = \int \rr \rob{t'} \phir \rob{t,t'}  dt'}_{\text{Measured Signal}} \ \xrightarrow{{{\textsf{Sampling}}}} \ \underbrace{\gr \sqb{n} = {\left. {\gr\left( t \right)} \right|_{t = nT}}}_{\text{Measured Samples}}
$}
\caption{Block diagram for ToF image formation process. The goal is to estimate $\sr$ from $\{\gr \sqb{n}\}_{n\in\id{N}}$. }
\label{fig:BlkTRI}
\end{figure*}

\bpara{Notation.} The set of integer, real, and complex-valued numbers are denoted by $ \mathbb{Z}, \mathbb{R}$ and $\mathbb{C}$, respectively. 
The set of $N$ contiguous integers is denoted by $\id{N} = \{0,\cdots, N-1\}, N\in \mathbb{Z}^{+}$. 
Continuous functions are written as $g\rob{t}, t\in \mathbb{R}$; their discrete counterparts are represented by 
$g\left[ n \right] = {\left. {g\left( t \right)} \right|_{t = nT}}$,
$n\in \mathbb{Z}$ where $T  > 0$ takes the role of sampling period. Vectors and matrices are written in bold lowercase and uppercase fonts, such as $\mathbf{g} = [g[0],\cdots,g[N-1]]^{\transp} \iR^{N}$ and $\mat{G} = [g_{n,m}]_{n\in\id{N}}^{m\in\id{M}} \iR^{N\times M}$.
The $\Lp{p}\rob{\R}$ space equipped with the $p$-norm or $\normt{\cdot}{p}{\R}$ is the standard Lebesgue space. For instance, $\Lp{1}$ and $\Lp{2}$ denote the space of absolute and square-integrable functions, respectively. Spaces associated with sequences are denoted by $\ell_{p}$. 
The max-norm $(\Lp{\infty})$ of a function is defined as, $\norm{g}_{\infty} = \inf \{c_0 \geqslant 0: \left|g\rob{t}\right| \leqslant c_0 \}$; for sequences, we use, $\norm{g}_{\infty} = \max_{n} \left| g \sqb{n} \right|$.
The $\Lp{2}$-norm of a function is defined as, $\norm{g}_{2} = \sqrt{\int \left|g\rob{t}\right|^{2} dt}$ while for sequences, we have, $\norm{g}_{2} = \sqrt{\sum_{n =0}^{N-1} \left| g \sqb{n} \right|^{2}}$.
The $\lp{0}$-norm of a sequence denotes the cardinality (\ie the number of non-zero entries) of the sequence. 
The inner-product of two functions $f,g\in \Lp{2}$ is defined as, $\inner{f}{g} = \int f(t)g^*(t) dt$ while for sequences, we have, $\inner{f}{g} = \sum_{n =0}^{N-1} f[n] g^*[n]$. 
The vector space of polynomials with complex coefficients and degrees less than or equal to $K$ is denoted by $P_{K}$, for instance, $\QMs (z) = \sum\nolimits_{k=0}^{K} h_{k} z^{k} \in P_{K}$. 
The $N$-order derivative of a function is denoted by $\df{N}{g}{t} \left( t \right)$. The space of first-order continuously differentiable, real-valued functions is denoted by $C\rob{\R}$. For sequences,  the first-order finite difference is denoted by $(\Delta g)\sqb{n} = g[n+1]- g\sqb{n}$.
For any exponential type functions $|g(t)| \leqslant Ae^{B|t|}$, its Laplace Transform is defined by $\Lap{g}{s} = \int_{0}^{\infty} g(t)e^{-st} dt, s\iC$. 
For any function $g\in \Lp{1}$, its Fourier Transform is defined by $\widehat{g} (\omega)  = \int g\rob{t} e^{-\jmath \omega t} dt$. For sequences, the Discrete Fourier Transform (DFT) of a sequence $\mat{g}\in \ell_{1}$ is denoted by $\widehat{g}[m] = \sum\nolimits_{n=0}^{N-1} g\sqb{n} e^{-\jmath \frac{2\pi n }{N}m}$. 
Let $\wmat{N}{M}$ and $\vmat{N}{M}$ denote the $N\times M$ 
\rgn{Vandermonde matrices} 
$\wmat{N}{M}=\bigl[ \zm{N}{-n\cdot m}  \bigr]_{n\in\id{N}}^{m\in\id{M}},\vmat{N}{M}=\frac{1}{N}\bigl[ \zm{N}{n\cdot m}  \bigr]_{n\in\id{N}}^{m\in\id{M}}, \ \zm{N}{n} = e^{\jmath\frac{2\pi n}{N}}$. The DFT of $\mathbf{g}$ can be expressed as $\widehat{\mathbf{g}} = \wmat{N}{N} \mathbf{g}$. 
The Gram matrix of $\mat{G}$ is defined as $\gram{\mat{G}} = \mat{G}^{\Hermit} \mat{G}$. 
The short hand notation for diagonal matrices is given by $\dk \rob{\mat{h}}$ with $\sqb{\dk \rob{\mat{h}}}_{k,k} = \sqb{\mat{h}}_{k\in\id{K}}$. 
The mean-squared error (MSE) between $\mat{x}, \mat{y} \iR^N$ is given by $\mse{x}{y} = \frac{1}{N}\sum\nolimits_{n=0}^{N-1} \left|x\sqb{n} -y\sqb{n} \right|^{2}$.

\section{Image Formation}
\label{sec:formation}

Let $\mathbf{r} = [x,y]^{\top}$ denote a point in the Cartesian coordinate where $(\cdot)^{\top}$ is the transpose operation. ToF sensors are active imaging systems that probe the \ttd scene of interest with some time-localized kernel \cite{Velten:2012:J,Velten:2016:J,Shin:2016:J,Satat:2015:J}, denoted by $\pr \rob{t}$ at a point $\mathbf{r}$. Based on the choice of kernel, ToF imaging can be classified as time-domain and frequency-domain ToF setups\footnote{While this paper is pivoted around time-localized kernels, to keep the exposition as general as possible, we will adhere to the generalized model described in our series of papers \cite{Bhandari:2016:C,Bhandari:2015:J,Bhandari:2016:J,Bhandari:2017:C,Bhandari:2017:Ca}. This model is general in the sense that it not only consolidates both time and frequency domain approaches but is also compatible with the broader theme of the ToF principle used in areas like terahertz \cite{RedoSanchez:2016:J}, ultrasound \cite{Tur:2011:J} and seismic imaging \cite{Claerbout:1973:J} as well as optical coherence tomography \cite{Seelamantula:2014:J,Blu:2002:C} and LIDAR \cite{Castorena:2015:J}. }. 

The emitted signal $\pr \rob{t}$ interacts with the \ttd scene characterized by the spatio-temporal scene response function (SRF) $\sr \rob{t,t'}$. In the case of multiple reflections (see \fig{fig:LIF}), the SRF is given by,
\begin{equation}
\label{eq:SRF}
\sr \rob{t,t'} = \sum\nolimits_{k = 0}^{K - 1} {\gammar \sqb{k}\delta  \rob{t - t'- \taur\sqb{k}} }
\end{equation}
where $\delta(\cdot)$ is a Dirac distribution and $\{\gammar \sqb{k},\taur\sqb{k}\}_{k\in\id{K}}$ are the corresponding reflectivities and time-delays ($\taur\sqb{k} = 2 {d _{\mathbf{r}}}\sqb{k}/c$) induced by $K$ light paths at point $\mathbf{r}$.\footnote{Note the SRF may be described more generally where $\sr\rob{t,t'} $ is a Green's function of some partial differential equation \cite{Bhandari:2015:J} or simply a transfer function. For instance, in the scenario of fluorescence lifetime imaging, $\sr\rob{t,t'}  = \sr^{\mbox{\tiny Depth}} \rob{t,t'} +\sr^{\mbox{\tiny Decay}}\rob{t,t'}$ for which $\sr^{\mbox{\tiny Depth}}\rob{t,t'}$ is defined in \eqref{eq:SRF} and represents a delay of $\taur\sqb{k}$ due to the fluorescent sample's placement at depth ${d _{\mathbf{r}}}\sqb{k}$ meters from the sensor and, $\sr^{\mbox{\tiny Decay}}\rob{t,t'} = \alpha_{\mathbf{r}} \mbox{exp} \left(-\rob{t-t'-\taur}/{\lambda}\right) h (t-t'-\taur)$ where $\alpha_{\mathbf{r}} $ and $\lambda$ are emission light strength and lifetime of the fluorescent sample, respectively and $h\rob{t}$ is the Heaviside step function. }

In several practical scenarios, the SRF can be written as a shift-invariant function, 
$
\sr\rob{t,t'} = \sr\rob{t-t'},
$
which reduces the measurements to a convolution format. This is because, the interplay of the emitted signal with the \ttd scene results in the reflected signal given by $\rr\rob{t} = \int \pr\rob{t'} \sr\rob{t,t'}  dt'$ resulting in convolution, 
$
\rr\rob{t} =   \rob{\pr \ast \sr}  \rob{t}.
$
The reflected signal is measured at the ToF sensor through its electro-optical architecture, which is described by its instrument response function (IRF), denoted by $\phir\rob{t,t'}$\footnote{In the absence of time-resolved perspective or steady-state scene assumption, the IRF is equivalent to the spatial point spread function.}. This leads to the continuous-time measurements defined by
$
\gr\rob{t} = \int \rr \rob{t'} \phir \rob{t,t'} dt'.
$
Again, in practice, the IRF $\phir \rob{t,t'}$ is typically shift-invariant, \ie $ \phir \rob{t,t'} = \phir \rob{t-t'}$. Consequently, the measurements impinging on the ToF imaging sensor simplify to,  
$
\gr\rob{t} = \rob{\rr \ast \phir}  \rob{t},
$
which can be further simplified as
\begin{equation}
\label{eq:ToF measurements}
\gr\left( t \right) = \left( {s_{\mathbf{r}}} \ast \fr \right) (t)\quad  \mbox{and} \quad \fr (t) \DE  \rob{\pr  \ast \phir} (t)
\end{equation}
where $\fr$ can be interpreted as the kernel in the spirit of \SR \cite{Candes:2013:J} or signal processing perspective. Finally, uniform sampling of the continuous-time input results in per-pixel digital measurements,
\begin{equation}
\label{eq:filtered samples}
\gr \sqb{n} = {\left. {\gr\left( t \right)} \right|_{t = nT}}=   \sum\nolimits_{k = 0}^{K - 1} {\gammar\sqb{k}\fr \left( {nT - \taur\sqb{k}} \right)}
\end{equation}
where $T>0$ is the sampling step and usually of the magnitude of $60$ to $100$ picoseconds (see \secref{sec:exp}). A global view of the image formation process is described as a mathematical block diagram in~\fig{fig:BlkTRI}. 

\bpara{Goal:} Starting with $\{\gr\sqb{n}\}_{n\in\id{N}}$ \emph{only}, our goal is to recover $\sr \rob{t}$ and $\fr(t)$ from possibly imperfect and distorted measurements. 

\begin{figure*}[!t]
\centering
\includegraphics[width=\columnwidth]{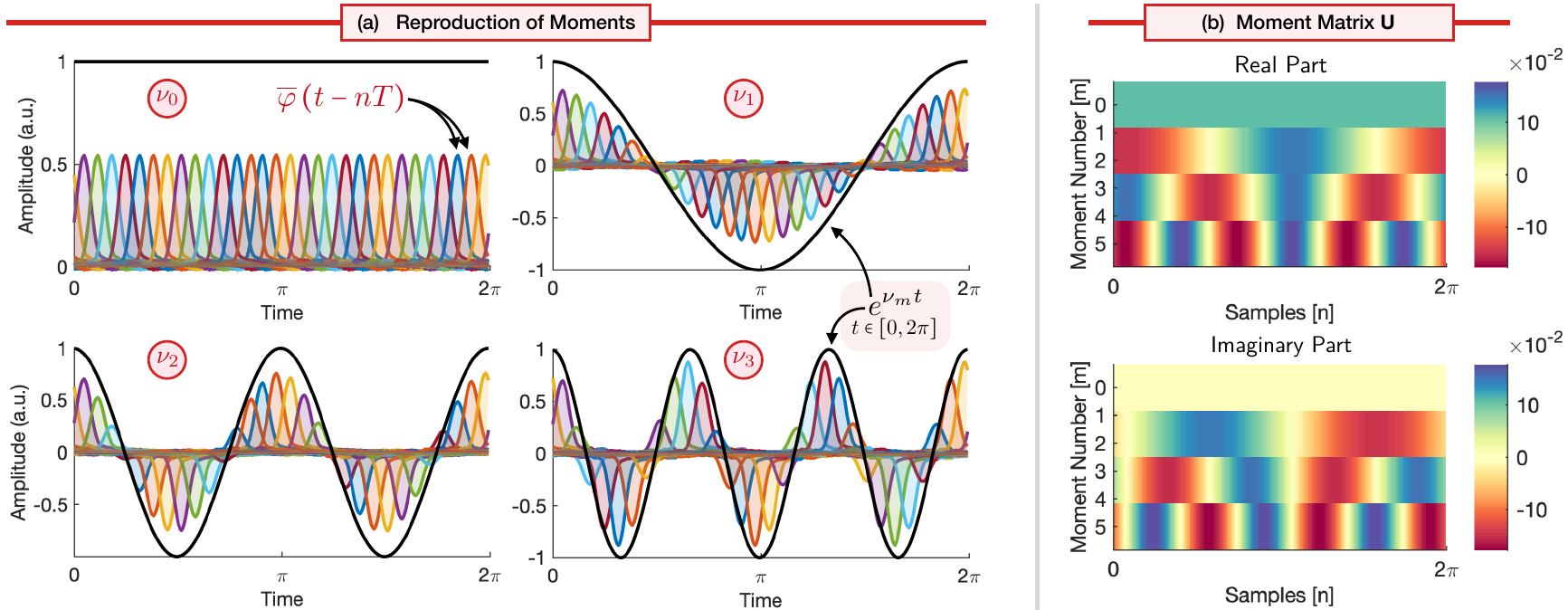}	
\caption{Using the kernel $\fr$ in \fig{fig:ToFkernel}~(a), we reproduce real part of exponentials ${e^{{\nu_m}t}}$ in (a) $m = 0,\ldots,3$. (b) Visualization of the moment coefficient matrix $\U$ in \eqref{eq:C formula}.}
\label{fig:exponential reproduction kernel}
\end{figure*}

\section{Super-Resolved ToF Imaging with Strang-Fix Properties}
\label{sec:SF}

Due to the shift-invariant property of the SRF and IRF, the ToF measurements depend on the kernel $\fr (t) \EQc{eq:ToF measurements} \left(p_{\mathbf{r}}  \ast \psi_{\mathbf{r}}\right) (t)$. The literature around the topic of ToF imaging is largely focused on the data model with discrete sparsity so that sparse recovery techniques can be leveraged \cite{Kadambi:2013:J,Toole:2017:C,Heide:2013:J,Qiao:2015:J}. However, such approaches artificially constrain delays on a discrete grid \ie $\{\taur\sqb{k}\}_{k\in\id{K}} \subseteq T\mathbb{Z}$, which is an artifact of the data model. We believe that this is the reason why \SR methods have not been widely reported in the ToF literature. 

In developing a blind ToF imaging strategy, an effective starting point involves abstracting properties of $\fr (t)$ \eqref{eq:ToF measurements} common across various ToF modalities \cite{Pellegrini:2000:J,Bhandari:2016:C,Bhandari:2015:J,Bhandari:2016:J,Bhandari:2017:C,Bhandari:2017:Ca,Kadambi:2013:J,Toole:2017:C,Heide:2013:J}, such as lock-in sensors and TCSPC systems. Our observation, based on various experimentally calibrated kernels used in ToF imaging \cite{Pellegrini:2000:J,Bhandari:2016:C,Bhandari:2015:J,Bhandari:2016:J,Bhandari:2017:C,Bhandari:2017:Ca,Kadambi:2013:J,Toole:2017:C,Heide:2013:J}, indicates that these kernels exhibit shared characteristics. By identifying and abstracting these common mathematical properties, we can pave the way for blind ToF imaging strategies. Specifically, our analysis of the literature highlights these shared features, setting the foundation for our approach.
\begin{enumerate}[leftmargin=*, label = $\bullet$] 
\item \textit{Time-localization}: $\fr (t) = 0,\; t \not\in \mathcal{D} \subset \mathbb{R}$ where $\mathcal{D}$ is finite, contiguous interval in the time-domain.  
\item \textit{Smoothness}: $\fr \in C\rob{\R}$ is continuously differentiable.
\end{enumerate}

\bpara{Towards Kernels with Strang-Fix Properties.} Given the nature of ToF kernels, a weighted linear combination of the kernel reproduces the prototype function,
\[
{t^{h}e^{ - \frac{{\nu _m} t}{T}}}, \quad h\in\N \mbox{ and } \nu_m \in \C
\]
and hence, implicit parameterization in terms of ${t^{h}e^{ - \frac{{\nu _m} t}{T}}}$ captures the features of the kernel in a \emph{flexible yet compressible manner}. As we shall see shortly, this compressibility proves to be advantageous in our blind ToF context. Similar properties have been widely studied in the wavelet and approximation theory, where they arise in the context of \emph{Strang-Fix Conditions} \cite{Strang:2011:B,Urigen:2013:J}. In particular, such kernels satisfy the property, 
\begin{equation}
\label{eq:reproducing kernel}
{t^{h}e^{ - \frac{{\nu _m} t}{T}}} = \sum\nolimits_{n \in \mathbb{Z}} {{\mu _{m,n,h}}\bpr  \left( {t - nT} \right)} ,\; m \in\id{M} \; \mbox{and} \; h\in\id{H}
\end{equation}
where $\bpr (t) =  \fr (-t)$ and $\{\mu _{m,n,h}\}_{n\in\Z}$ are the exponential polynomial reproducing coefficients. A visual illustration of \eqref{eq:reproducing kernel} is shown in \fig{fig:exponential reproduction kernel}. In other words, $\{t^{h}{e^{ - {\nu _m}\frac{t}{T}}}\}_{m\in\id{M}}$ belongs to a shift-invariant space spanned by shifts of $\fr$. For this to work, the kernel $\fr$ must satisfy what is known as the {generalized Strang-Fix conditions} \cite{Urigen:2013:J}:
\begin{equation}
\label{eq:Strang-Fix conditions}
{\left. {\df{h}{\Lap{\fr}{\nu}}{\nu} } \right|_{\nu = \nu_m}}   \neq 0 \ \ \mbox{and} \ \ {\left. {\df{h}{\Lap{\fr}{\nu}}{\nu} } \right|_{\nu = \nu_m + \jmath\frac{2\pi l}{T}}}   = 0, \; l \in \mathbb{Z}/\{0\}, \; m \in\id{M} \; \mbox{and} \; h\in\id{H}
\end{equation}
where $\Lap{\fr}{s}$ denotes the Laplace Transform of $\fr$ and $M$ in \eqref{eq:reproducing kernel} is the order of the kernel.

The main advantage of the parametrization in \eqref{eq:reproducing kernel} is that the continuous-time unknown kernel ($\fr\rob{t}$) can be expressed in a low-dimensional, discrete representation, $\{\mu _{m,n,h}\}_{n\in\Z}$ which makes the formulation of our optimization method efficient and tractable.

\bpara{Perfect Recovery of SRF with Known Kernel.} 
Before developing blind ToF imaging methodology, we consider the intermediate step where the kernel is assumed to be known. In this setting, we consider the following two fundamental questions:
\begin{enumerate*}[leftmargin = *, label = \uline{\arabic*})]
\item What is the mathematical criterion for perfect recovery of the \srf?
\item Is there a constructive algorithm for recovery?
\end{enumerate*}
In what follows, we will demonstrate a \srf recovery method that relies on transform domain characterization of spikes and leverages spectral estimation. This intermediate step is similar to previous works \cite{DeFigueiredo:1982:J,Candes:2013:J,Urigen:2013:J}.

\begin{proposition}
\label{thm:annihilation filter}
Let the kernel $\fr (t)=0, t \notin \mathcal{D} \subset \left[ 0,{{\tau }}\right)$ be known and further, let us assume that $\fr (t)$ satisfies Strang-Fix properties,
$
{e^{ - \frac{{\nu _m} t}{T}}} = \sum\nolimits_{n \in \mathbb{Z}} {{\mu _{m,n}}\bpr  \left( {t - nT} \right)} ,\; m \in\id{M}.
$
Given the measurements defined as,
$
\gr \sqb{n} = \rob{s_{\mathbf{r}} \ast \fr} \rob{nT} \equiv   \sum\nolimits_{k = 0}^{K - 1} {\gammar\sqb{k}\fr \left( {nT - \taur\sqb{k}} \right)}
$
then, the \srf $\sr \rob{t}  = \sum\nolimits_{k = 0}^{K - 1} {\gammar \sqb{k}\delta  \rob{t - \taur\sqb{k}} }$ can be recovered from $N\geq 2K$ samples.
\end{proposition}	
\begin{proof}
As detailed out in \eqref{eq:C formula}, \lemmaref{lemma:moment decomposition}, the exponential reproducing coefficients $\{\mu _{m,n}\}_{m\in\id{M}}^{n\in\id{N}}$ can be computed given $\fr$. With coefficients $\{\mu _{m,n}\}_{m\in\id{M}}^{n\in\id{N}}$ known, let us define {exponential moments}, 
\begin{align}
\label{eq:kernel reproduction}
\yr[m]&=\sum\nolimits_{n=0}^{N-1} \mu_{m,n}\gr [n]
\EQc{eq:filtered samples} \sum\nolimits_{n =0}^{N-1} \mu _{m,n}\sum\nolimits_{k = 0}^{K - 1} {{\Gamma _{\mathbf{r}}}\left[ k \right]\fr \left( {nT - {\tau _{\mathbf{r}}}\left[ k \right]} \right)}  \nonumber \\
&= \sum\nolimits_{k=0}^{K-1} {\Gamma _{\mathbf{r}}}\left[ k \right] {\sum\nolimits_{n =0}^{N-1} \mu _{m,n} \bpr \left( { {\tau _{\mathbf{r}}}\left[ k \right]- nT} \right) }
\eqr{eq:reproducing kernel}\sum\nolimits_{k=0}^{K-1} \gammar[k] {e^{ \frac{- {\nu _m}}{T}{\taur}\left[ k \right]}}
\end{align}
which is a finite sum of $K$-complex exponentials. Let $\nu _m = \jmath \frac{2m\pi }{N}, m\in\id{M}$. The \emph{unknown} frequencies in \eqref{eq:kernel reproduction} can be found by using Prony's method as follows. Let $h_{\mathbf{r}} \sqb{m}$ be the filter with $z$-transform,
\begin{equation}
\label{eq:h filter}
\H(z)  = \sum\nolimits_{m=0}^{K} h \sqb{m} z^{-m} = \prod\nolimits_{k=0}^{K-1} (1 - u_k z^{-1}), \; u_k = e^{ \frac{- \jmath 2\pi }{\tau}\taur\sqb{k}} \; \mbox{and} \; \H\in P_{K}
\end{equation}
where $\{u_k\}_{k\in\id{K}}$ are the roots of $\H$ since $\H(u_k) = 0, k\in\id{K}$. 
Then, $h \sqb{m}$ annihilates $\yr\sqb{m}$ or, 
\begin{equation}
\label{eq:annihilation}
(h \ast \yr)\sqb{m} = \sum\nolimits_{l=0}^{K} h [l]  \yr[m-l] 
= \sum\nolimits_{k=0}^{K-1} \gammar[k]
(\sum\nolimits_{l=0}^{K} h [l] u_k^{-l})
u_k^{m}  = 0.
\end{equation}
The annihilation filter $\{h[m]\}_{m\in\id{K+1}}$ can be found by solving a system of linear equations in \eqref{eq:annihilation} while $\{\taur[k]\}_{k\in\id{K}}$ can be obtained by computing the zeros $u_k$ of the polynomial $\H$ constructed in \eqref{eq:h filter}. The amplitudes $\{\gammar[k]\}_{k\in\id{K}}$ can be computed via least-squares since both $\{\yr [m]\}_{m\in\id{M}}$ and $\{\taur[k]\}_{k\in\id{K}}$ are already known. The problem can be solved as soon as there are at least many equations as unknowns; \ie $N \geqslant 2K$ samples for estimating $2K$ unknowns $\{\gammar[k], \taur[k]\}_{k\in \id{K}}$. 
\end{proof}

Until this point, we have seen that kernels that satisfy Strang-Fix properties can lead to sparse \SR recovery. That said, it still remains to justify why ToF kernels can reproduce exponentials and what is a numerical method to compute the exponential reproducing coefficients $\{\mu _{m,n}\}_{m\in\id{M}}^{n\in\id{N}}$.

\bpara{ToF Kernels Satisfy Strang-Fix Conditions.}
\rg{
In this part, we demonstrate that the common features of the ToF kernel, \ie, \emph{time-localization} and \emph{smoothness}, mathematically results in a decaying Fourier spectrum, and thus the Strang-Fix conditions in \eqref{eq:Strang-Fix conditions} holds approximately. More specifically, we have,
\begin{enumerate}[leftmargin=*, label = ---$\bullet$] 
\item Time-localization: Since the kernel satisfies $\fr (t) = 0,\; t \not\in \mathcal{D} \subset \mathbb{R}$, we can express $\fr (t)$ as
\begin{equation}
\label{eq:Fourier series}
\fr(t) = \sum\nolimits_{m \in \mathbb{Z}} \bp[m] \e^{\jmath \frac{2m\pi t}{\tau}}, \;  \bp\sqb{m} = \frac{1}{\tau}\int_{0}^{\tau} {\fr(t)}  \e^{-\jmath \frac{2m\pi t}{\tau}} dt, 
\; \mbox{and} \;
\tau > \max_{k,\mathbf{r}} |\taur [k]|.
\end{equation}
\item Smoothness: Let $\fr \in C\rob{\R}$ be a time-localized kernel with $\fr\rob{0} = \fr\rob{\tau}$\footnote{Without loss of any generality, we use the boundary condition to simplify the result in \eqref{eq:decay} which is a standard result in Fourier analysis. }, then its Fourier series coefficients decay as frequency increases, \ie 
\begin{equation}
\label{eq:decay}
\abs{\bp\left[m\right]}  \leqslant \frac{\tau}{\left| 2m \pi\right|} \norm{ \df{1}{\fr}{t}  \left( t \right)}_{\infty}, \; m\in \mathbb{Z}/\{0\}.
\end{equation}
\end{enumerate}
A consequence of these two properties is that: $\{{\bp}[m]\}_{m\iZ}$ contract to zero as $\abs{m}$ increases, \ie, 
\begin{equation}
\label{eq:approxSF}
\forall \epsilon >0,  \exists M \in\N^{+} \Rightarrow \abs{\bp\left[m\right]} < \epsilon, \abs{m} > M
\end{equation}
and hence $\fr$ is \rgnn{well} approximated by a finite number of Fourier series coefficients.
\rgnn{That is to say, \eqref{eq:approxSF} implicitly mimics Strang-Fix conditions in \eqref{eq:Strang-Fix conditions} ($H=0$) since for every $\abs{m} > M$, we have ${|\bp\left[m\right]|} \approxeq 0 \implies \abs{\Lap{\fr}{\jmath \frac{2m\pi }{\tau}}} \approxeq 0$, with an error bounded by $\tfrac{\tau}{| 2M \pi |} ||{ \df{1}{\fr}{t}  \left( t \right)}||_{\infty}$.} Hence, $\fr$ approximately satisfies the Strang-Fix conditions in \eqref{eq:Strang-Fix conditions}.}

\bpara{Computation of Exponential Reproducing Coefficients.} Given the kernel $\fr (t) = (p_{\mathbf{r}}  \ast \psi_{\mathbf{r}}) (t), t\in\left[0, \tau\right)$, we now present an efficient method to compute exponential reproducing coefficients. 
\begin{lemma}
\label{lemma:moment decomposition}
Let the kernel $\fr (t)=0, t \notin \mathcal{D} \subset \left[ 0,\tau \right)$ be known and further, let us assume that $\fr (t)$ satisfies Strang-Fix properties,
$
{e^{ - \frac{{\nu _m} t}{T}}} = \sum\nolimits_{n \in \id{N}} {{\mu _{m,n}}\bpr  \left( {t - nT} \right)} ,\; m \in\id{M}.
$
Let $\nu _m = \jmath \frac{2m\pi }{N}, m\in\id{M}$. Then, the coefficients $\{\mu _{m,n}\}_{n\in\id{N}}^{m\in\id{M}}$ are given by
\begin{equation}
\label{eq:C formula}
\ {\U} = \frac{1}{N}\dk^{-1} \rob{\mbp} \wmat{M}{N}, \; \U = \sqb{\mu _{m,n}}_{m\in\id{M}}^{n\in\id{N}} 
\; \mbox{and} \; \tau=NT.
\end{equation}
\end{lemma}
\begin{proof}

From the exponential reproduction properties in \eqref{eq:reproducing kernel}, we obtain that,
\begin{align}
\label{eq:C property}
\int_{0}^{\tau} \sum\limits_{n=0}^{N-1} {\mu _{m,n}}  \bpr  ( {t - nT} ) {e^{  \frac{{\nu _l} t}{T}}}  dt 
\eqr{eq:reproducing kernel} \int_{0}^{NT} {e^{ ({\nu _l}- {\nu _m})\frac{t}{T}}} dt 
=  \tau \delta \sqb{l-m}
\end{align}
where $\delta \sqb{\cdot}$ is the Kronecker delta \rgnn{sequence}. Since $\nu _m = \jmath \frac{2m\pi }{N}, m\in\id{M}$, then $\fr$ satisfies that,
\begin{equation}
\label{eq:period}
\int_{0}^{\tau} \bpr  \left( {t - nT} \right) {e^{  \frac{{\nu _l}t}{T}}}  dt 
= \tau \bp\sqb{l}   \e^{\nu_l n}.   
\end{equation}
Hence, combining \eqref{eq:C property} and \eqref{eq:period}, we have that  
\begin{equation}
\label{eq:moment relation}
\int_{0}^{\tau} \sum\limits_{n=0}^{N-1} {\mu _{m,n}}  \bpr  ( {t - nT} ) {e^{  \frac{{\nu _l} t}{T}}}  dt  = \tau \sum\nolimits_{n =0}^{N-1} {\mu _{m,n}}  \bp\sqb{l}   \e^{\nu_l n}  =   \tau \delta \sqb{l-m}, \; l,m\in\id{M}
\end{equation}
which can be algebraically rewritten in matrix form as,
$
{\U} \vmat{N}{M}  \dk \rob{\mbp}  = \frac{1}{N}\I
$
where $\I\is{R}^{M\times M}$ is the identity matrix. 
By simplification, we eventually have that,
\[
\renewcommand\matscale{.55}
\matbox{7}{11}{M}{N}{\U} \ =\frac{1}{N} \ 
\matbox{7}{7}{M}{M}{\dk^{-1} \rob{\mbp}} \ \ 
\matbox{7}{11}{M}{N}{\wmat{M}{N}}
\]
\begin{equation}
{\U} = \frac{1}{N}\dk^{-1} \rob{\mbp} \wmat{M}{N} \quad \mbox{where} \quad \sqb{  \dk \rob{\mbp} }_{m,m} = \bp\sqb{m}.
\end{equation}
\end{proof}

\section{Blind Super-Resolved ToF Imaging}
\label{sec:BSD}

\begin{figure}[!tb]
\centering
\scalebox{0.8}{\tikzset{every picture/.style={line width=0.75pt}} 

\begin{tikzpicture}[x=0.75pt,y=0.75pt,yscale=-1,xscale=1]

\draw  [fill={rgb, 255:red, 155; green, 155; blue, 155 }  ,fill opacity=0.1 ][line width=1.5]  (170,120) .. controls (170,114.48) and (174.48,110) .. (180,110) -- (510,110) .. controls (515.52,110) and (520,114.48) .. (520,120) -- (520,150) .. controls (520,155.52) and (515.52,160) .. (510,160) -- (180,160) .. controls (174.48,160) and (170,155.52) .. (170,150) -- cycle ;
\draw [line width=1.5]    (340,160) -- (340,186) ;
\draw [shift={(340,190)}, rotate = 270] [fill={rgb, 255:red, 0; green, 0; blue, 0 }  ][line width=0.08]  [draw opacity=0] (11.61,-5.58) -- (0,0) -- (11.61,5.58) -- cycle    ;
\draw [line width=1.5]    (340,240) -- (340,279) ;
\draw [shift={(340,283)}, rotate = 270] [fill={rgb, 255:red, 0; green, 0; blue, 0 }  ][line width=0.08]  [draw opacity=0] (11.61,-5.58) -- (0,0) -- (11.61,5.58) -- cycle    ;
\draw  [fill={rgb, 255:red, 155; green, 155; blue, 155 }  ,fill opacity=0.1 ][line width=1.5]  (180,200) .. controls (180,194.48) and (184.48,190) .. (190,190) -- (500,190) .. controls (505.52,190) and (510,194.48) .. (510,200) -- (510,230) .. controls (510,235.52) and (505.52,240) .. (500,240) -- (190,240) .. controls (184.48,240) and (180,235.52) .. (180,230) -- cycle ;
\draw    (130,540) -- (130,580) ;
\draw    (470,550) -- (470,590) ;

\draw  [color={rgb, 255:red, 0; green, 0; blue, 255 }  ,draw opacity=1 ]  (518.5,268) -- (626.5,268) -- (626.5,310) -- (518.5,310) -- cycle (515.5,265) -- (629.5,265) -- (629.5,313) -- (515.5,313) -- cycle ;
\draw (572.5,289) node  [font=\small] [align=left] {\begin{minipage}[lt]{70.58pt}\setlength\topsep{0pt}
\begin{center}
Sub-Problem P2\\Kernel Recovery
\end{center}

\end{minipage}};
\draw  [color={rgb, 255:red, 0; green, 0; blue, 255 }  ,draw opacity=1 ]  (27.5,268) -- (136.5,268) -- (136.5,310) -- (27.5,310) -- cycle (24.5,265) -- (139.5,265) -- (139.5,313) -- (24.5,313) -- cycle ;
\draw (82,289) node  [font=\small] [align=left] {\begin{minipage}[lt]{71.09pt}\setlength\topsep{0pt}
\begin{center}
Sub-Problem P1\\Spike Estimation
\end{center}

\end{minipage}};
\draw  [color={rgb, 255:red, 200; green, 0; blue, 0 }  ,draw opacity=1 ][fill={rgb, 255:red, 245; green, 235; blue, 230 }  ,fill opacity=0.52 ][line width=1.5]   (340, 305) circle [x radius= 71.42, y radius= 20.51]   ;
\draw (340,305) node  [font=\large]  {$
\gr\sqb{n}, n\in\id{N}
$};
\draw    (519.5, 351.5) circle [x radius= 45.25, y radius= 20.51]   (519.5, 351.5) circle [x radius= 48.25, y radius= 23.51]  ;
\draw (519.5,351.5) node  [font=\large,color={rgb, 255:red, 190; green, 0; blue, 0 }  ,opacity=1 ]  {$
\fr^{[i]} \sqb{n} 
$};
\draw  [color={rgb, 255:red, 0; green, 0; blue, 0 }  ,draw opacity=1 ]  (133, 344.5) circle [x radius= 38.18, y radius= 20.51]   (133, 344.5) circle [x radius= 41.18, y radius= 23.51]  ;
\draw (133,344.5) node  [font=\large,color={rgb, 255:red, 190; green, 0; blue, 0 }  ,opacity=1 ]  {$
\qr^{[i]} \sqb{n} 
$};
\draw  [fill={rgb, 255:red, 155; green, 155; blue, 155 }  ,fill opacity=0.1 ]  (295,339.5) -- (390,339.5) -- (390,362.5) -- (295,362.5) -- cycle  ;
\draw (342.5,351) node  [font=\small] [align=left] {{\small \textcolor[rgb]{0.75,0,0}{{\fontfamily{helvet}\selectfont  Measurements }}}};
\draw (431,418.4) node [anchor=north west][inner sep=0.75pt]  [font=\small]  {$
\mphi^{[i]} = \rob{\gram{\Toep{\mqr}} }^{-1}  \Toep{\mqr}^{\Hermit} \mgr
$};
\draw (48,402.4) node [anchor=north west][inner sep=0.75pt]  [font=\small]  {$ \begin{array}{l}
\qr[n] = {\PMs (\zm{N}{n})}/{\QMs (\zm{N}{n})}, \\
\PMs \in P_{K-1}, \QMs \in P_{K}
\end{array}$};
\draw  [draw opacity=0]  (163.5,112.1) -- (529.5,112.1) -- (529.5,134.1) -- (163.5,134.1) -- cycle  ;
\draw (346.5,123.1) node  [font=\large,color={rgb, 255:red, 200; green, 0; blue, 0 }  ,opacity=1 ] [align=left] {\begin{minipage}[lt]{246.27pt}\setlength\topsep{0pt}
\begin{center}
Continuous-Time ToF Sensing Model (\secref{sec:SF})
\end{center}

\end{minipage}};
\draw (347.5,143) node  [font=\large]  {$
\gr \sqb{n} =   \sum\nolimits_{k = 0}^{K - 1} {\gammar\sqb{k}\fr \left( {nT - \taur\sqb{k}} \right)}
$};
\draw  [draw opacity=0]  (190.45,192.87) -- (496.45,192.87) -- (496.45,214.87) -- (190.45,214.87) -- cycle  ;
\draw (343.45,203.87) node  [font=\large] [align=left] {\begin{minipage}[lt]{205.66pt}\setlength\topsep{0pt}
\begin{center}
\textcolor[rgb]{0.78,0,0}{Discrete-Time Sensing Model (\secref{sec:BSD})}
\end{center}

\end{minipage}};
\draw (341.97,223.77) node  [font=\large]  {$
\gr \sqb{n} = \rob{\fr \circov \qr} \sqb{n} 
$};
\draw  [color={rgb, 255:red, 0; green, 0; blue, 0 }  ,draw opacity=1 ]  (131, 515.5) circle [x radius= 83.44, y radius= 21.92]   (131, 515.5) circle [x radius= 86.44, y radius= 24.92]  ;
\draw (131,515.5) node  [font=\large,color={rgb, 255:red, 190; green, 0; blue, 0 }  ,opacity=1 ]  {$\Aj ,\Bj ,\mur$};
\draw  [color={rgb, 255:red, 0; green, 0; blue, 0 }  ,draw opacity=1 ]  (469, 525.5) circle [x radius= 72.12, y radius= 21.92]   (469, 525.5) circle [x radius= 75.12, y radius= 24.92]  ;
\draw (469,525.5) node  [font=\large,color={rgb, 255:red, 190; green, 0; blue, 0 }  ,opacity=1 ]  {$
\p^{[j+1]},\q^{[j+1]}
$};
\draw (331,589.4) node [anchor=north west][inner sep=0.75pt]  [font=\small]  {
$ \begin{array}{l}
\{\p^{[j+1]},\q^{[j+1]}\}=\mathop{\rm arg\,min}\limits_{\p, \q}
\  \norm{\mur + \Aj \q - \Bj \p}_{2}^{2}  \\
\mbox{subject to}  \quad  \frac{1}{2\pi} \int_{0}^{2\pi}\rob{\QMs^{[0]}(e^{-\jmath\theta})}^{\ast}\QMs^{[j+1]}(e^{-\jmath\theta})d\theta=1 \notag
\end{array}$
};
\draw (11,577.4) node [anchor=north west][inner sep=0.75pt]  [font=\small]  {$ \begin{array}{l}
\Aj= \Toep{\fr} \Rj \dk (\mqr^{[0]}) \vmat{N}{K+1},\\
\Bj=  \Toep{\fr} \Rj \vmat{N}{K}, \\
\mur = \mgr - \Toep{\fr} \rob{\gram{\Toep{\fr}} }^{-1}  \Toep{\fr}^{\Hermit} \mgr, \\ 
\Rj = \rob{\dk(\vmat{N}{K+1} \q^{[j]})}^{-1}, \\ 
\mqr^{[0]} = \rob{\gram{\Toep{\fr}} }^{-1}  \Toep{\fr}^{\Hermit} \mgr
\end{array}$};
\draw    (375,279.69) .. controls (384.08,271.6) and (398.51,265.66) .. (413.43,265.66) .. controls (441.29,265.66) and (471.09,286.53) .. (502.66,328.79)(376.94,281.99) .. controls (385.41,274.42) and (399.17,268.66) .. (413.43,268.66) .. controls (440.65,268.66) and (469.54,289.47) .. (500.25,330.59) ;
\draw [shift={(369.38,286.3)}, rotate = 317.91] [color={rgb, 255:red, 0; green, 0; blue, 0 }  ][line width=0.75]    (10.93,-3.29) .. controls (6.95,-1.4) and (3.31,-0.3) .. (0,0) .. controls (3.31,0.3) and (6.95,1.4) .. (10.93,3.29)   ;
\draw    (474.51,372.27) .. controls (435.89,390.49) and (389.24,400.99) .. (340.61,400.99) .. controls (284.92,400.99) and (226.63,387.22) .. (165.74,359.65)(473.29,369.53) .. controls (435.02,387.59) and (388.8,397.99) .. (340.61,397.99) .. controls (285.32,397.99) and (227.44,384.29) .. (166.98,356.92) ;
\draw [shift={(482.81,366.77)}, rotate = 154.3] [color={rgb, 255:red, 0; green, 0; blue, 0 }  ][line width=0.75]    (10.93,-3.29) .. controls (6.95,-1.4) and (3.31,-0.3) .. (0,0) .. controls (3.31,0.3) and (6.95,1.4) .. (10.93,3.29)   ;
\draw    (155.29,315.97) .. controls (179.54,284.38) and (209.87,266.32) .. (242.36,266.32) .. controls (262,266.32) and (282.47,272.92) .. (303.75,286.19)(157.98,317.42) .. controls (181.18,287.17) and (210.65,269.32) .. (242.36,269.32) .. controls (261.51,269.32) and (281.43,275.81) .. (302.16,288.74) ;
\draw [shift={(151.33,323.45)}, rotate = 307.11] [color={rgb, 255:red, 0; green, 0; blue, 0 }  ][line width=0.75]    (10.93,-3.29) .. controls (6.95,-1.4) and (3.31,-0.3) .. (0,0) .. controls (3.31,0.3) and (6.95,1.4) .. (10.93,3.29)   ;
\draw    (133.33,398) -- (133.15,368.01) ;
\draw    (518.73,375) -- (517.46,414) ;
\draw    (134.27,451.04) -- (133.4,482.62)(131.27,450.96) -- (130.4,482.54) ;
\draw [shift={(131.68,490.58)}, rotate = 271.57] [color={rgb, 255:red, 0; green, 0; blue, 0 }  ][line width=0.75]    (10.93,-3.29) .. controls (6.95,-1.4) and (3.31,-0.3) .. (0,0) .. controls (3.31,0.3) and (6.95,1.4) .. (10.93,3.29)   ;
\draw    (167.54,491.69) .. controls (209.36,468.73) and (251.35,457.26) .. (293.5,457.26) .. controls (340.91,457.26) and (388.55,471.77) .. (431.36,497.85)(168.98,494.32) .. controls (210.32,471.62) and (251.83,460.26) .. (293.5,460.26) .. controls (340.4,460.26) and (387.51,474.65) .. (429.86,500.44) ;
\draw [shift={(436.96,502.95)}, rotate = 211.53] [color={rgb, 255:red, 0; green, 0; blue, 0 }  ][line width=0.75]    (10.93,-3.29) .. controls (6.95,-1.4) and (3.31,-0.3) .. (0,0) .. controls (3.31,0.3) and (6.95,1.4) .. (10.93,3.29)   ;
\draw    (409.35,541.97) .. controls (373.07,550.06) and (338.11,554.1) .. (304.48,554.1) .. controls (263.61,554.1) and (224.68,548.13) .. (193.27,537.95)(408.7,539.04) .. controls (372.64,547.08) and (337.9,551.1) .. (304.48,551.1) .. controls (263.93,551.1) and (225.31,545.18) .. (194.14,535.08) ;
\draw [shift={(187.16,534.45)}, rotate = 18.05] [color={rgb, 255:red, 0; green, 0; blue, 0 }  ][line width=0.75]    (10.93,-3.29) .. controls (6.95,-1.4) and (3.31,-0.3) .. (0,0) .. controls (3.31,0.3) and (6.95,1.4) .. (10.93,3.29)   ;

\end{tikzpicture}}  

\caption{Block diagram of the proposed blind ToF imaging methodology. In the absence of kernel calibration, our setting takes the form of \BSD problem, where we split it into two tractable sub-problems. 
$\PO$ that addresses recovery of $\qr[n]$ via continuous-time spike estimation (see \secref{subsec:spike}) and $\PT$ that solves for $\fr[n]$ utilizing least-squares fitting (see \secref{subsec:kernel}). 
}
\label{fig:BSD map}
\end{figure}
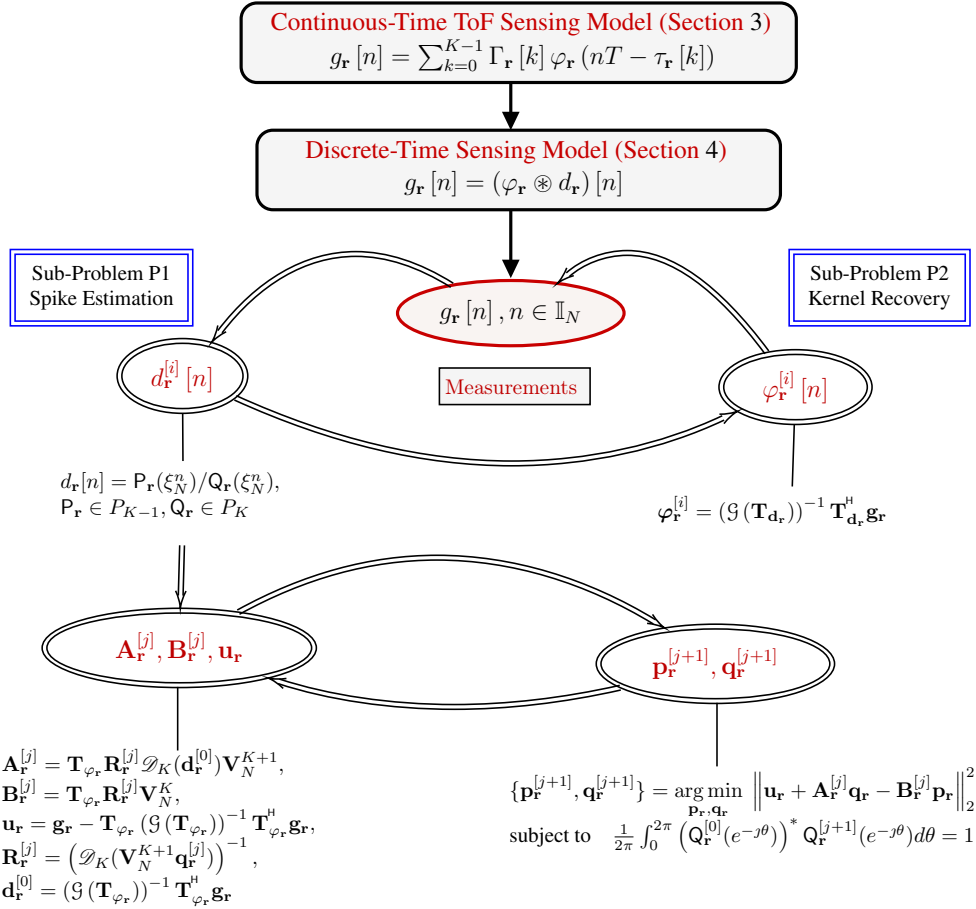

\rg{
In the intermediate scenario of a known kernel, we demonstrate that the \SR ToF imaging can be achieved by employing the exponential reproduction properties of the kernel.
In this section, we investigate the blind ToF imaging methodology that allows for adaptive estimation of both the \srf and kernel. 
In the absence of kernel calibration, our setup takes the shape of \BSD problem, which is generally challenging to solve due to its ill-posed formulation.
In order to tackle this problem, we opt for an alternating minimization framework that splits the \BSD into two tractable sub-problems, as described in \fig{fig:BSD map}.
}

\subsection{Sparse Model for ToF Measurements}

We first focus on a special subset of Strang-Fix kernels, \ie, bandlimited kernels. This scenario reveals an exact model that relates discrete ToF measurements $\{\gr\sqb{n}\}_{n\in\id{N}}$ to a continuous-time spike representation. When the bandlimitedness constraint is relaxed, we are left with an approximate model which, as shown via experiments, serves as a compelling match to real-world scenarios.
\begin{proposition}
\label{thm:signal model}
Let the kernel $\fr$ in \eqref{eq:Fourier series} be bandlimited, \ie, $\bp[m]= 0, \abs{m}>L$.  Then, 
\begin{align}
\gr \sqb{n} & = \rob{s_{\mathbf{r}} \ast \fr} \rob{nT} \equiv   \sum\nolimits_{k = 0}^{K - 1} {\gammar\sqb{k}\fr \left( {nT - \taur\sqb{k}} \right)} \notag    \\ 
& \equiv \rob{\fr \circov \qr} \sqb{n} \quad \mbox{where} \quad 
\qr\sqb{n} \DE \frac{1}{N}\sum\limits_{k=0}^{K-1}  \frac{{{\Gamma _{\mathbf{r}}}\left[ k \right]} \big(1 - (u_k)^N \big)  }{(1 - u_k \zm{N}{n}) }  = \frac{\PMs (\zm{N}{n})}{\QMs (\zm{N}{n})}
\label{eq:circ model}
\end{align}
and where $\circov$ denotes circular convolution, $\PMs \in P_{K-1}, \QMs \in P_{K}$, $u_k \EQc{eq:h filter} e^{ \frac{- \jmath 2\pi }{\tau}\taur\sqb{k}}$ and $\zm{N}{n} = e^{\jmath\frac{2\pi n}{N}}$.
\end{proposition}
\begin{figure}[!tb]
\centering
\includegraphics[width=0.7\columnwidth]{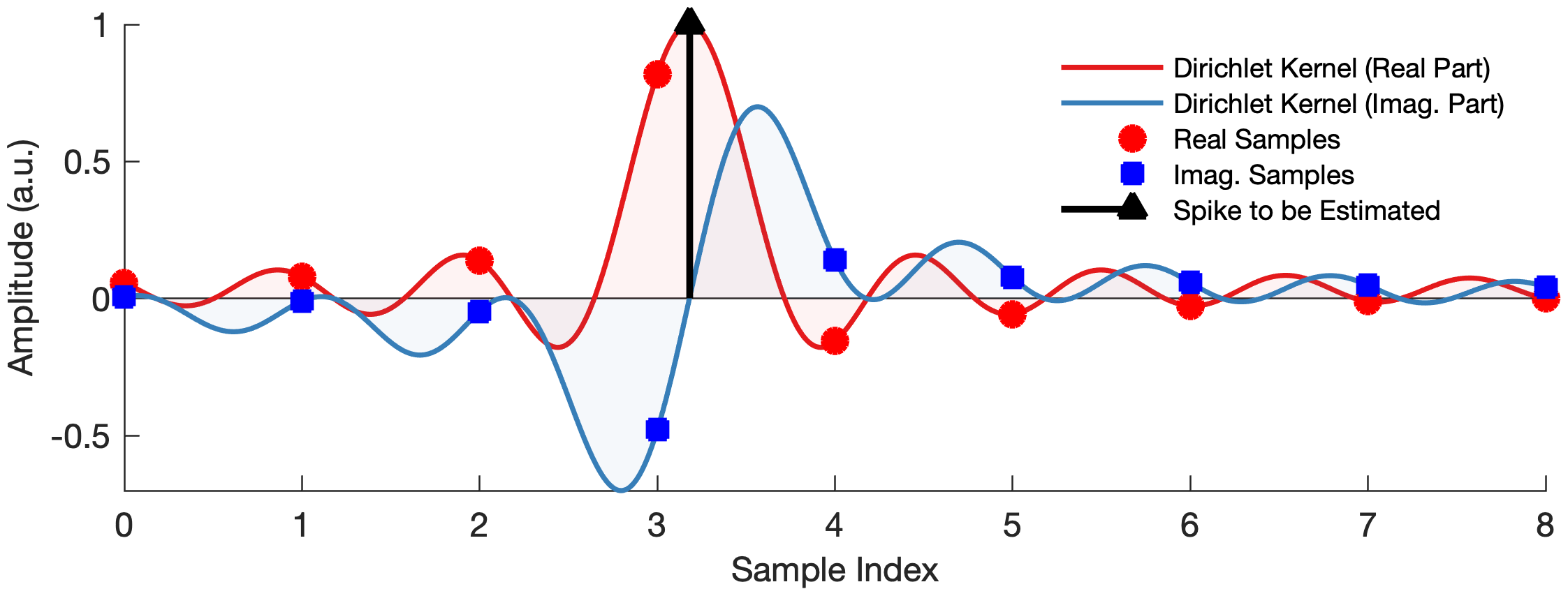} 		
\caption{Visualization of $\qr$ in \eqref{eq:circ model}. 
}
\label{fig:Dirichlet Model}
\end{figure}
\begin{proof}
Our starting point is the low-pass structure of the kernel $\fr$, resulting in
\begin{equation}
\label{eq:LP}
\fr \sqb{n}  = {\left. {\fr\left( t \right)} \right|_{t = nT}} = \inner{\fr(t)}{\fl{L}(t-nT)}, n\in\id{N}
\end{equation}
{where} $\fl{L}(t) = {\sin \rob{\rob{2L+1} \pi t/\tau}}/{\sin \rob{ \pi t/\tau}}$ is the low-pass filter.

Let $\hp$ denote the Discrete Fourier Transform (DFT) of $\fr$. We show that the DFT coefficients $\{\hp[l]\}_{l\in\id{M}}$ and its continuous counterpart $\{\bp[l]\}_{l\in\id{M}}$ are linearly dependent, since
\begin{align}
\label{eq:bh}
\hp[l] &= \sum\nolimits_{n =0}^{N-1} \fr [n] \zm{N}{-n\cdot l}  \EQc{eq:LP} \inner{\fr(t)}{\sum\nolimits_{n =0}^{N-1} \sum\nolimits_{|m|\leqslant L} e^{-\jmath \frac{2m\pi t}{\tau}}  e^{\jmath \frac{2n\pi (m-l)}{N}}} \notag \\
&=\inner{\fr(t)}{\sum\nolimits_{|m|\leqslant L}  e^{-\jmath \frac{2m\pi t}{\tau}} \delta \sqb{l-m}} = \inner{\fr(t)}{e^{-\jmath \frac{2l\pi t}{\tau}}} \eqr{eq:Fourier series} \tau \bp[l].
\end{align}
Similarly, we have 
$
\hgr[l] = \tau \bgr [l]
$
where $\bgr[m]= \frac{1}{\tau}\int_{0}^{\tau} {\gr(t)}  \e^{-\jmath \frac{2m\pi t}{\tau}} dt$. Moreover, notice that,
\begin{align*}
\gr(t) &= \sum\nolimits_{m \in \mathbb{Z}} \bgr[m] \e^{\jmath \frac{2m\pi t}{\tau}} 
= \sum\nolimits_{k = 0}^{K - 1} {\gammar\sqb{k}\fr \left( {t - \taur\sqb{k}} \right)}  \\
&= \sum\nolimits_{m \in \mathbb{Z}} (\bp[m] \sum\nolimits_{k = 0}^{K - 1} \gammar\sqb{k}  \e^{-\jmath \frac{2m\pi \taur[k]}{\tau}}) \e^{\jmath \frac{2m\pi t}{\tau}}.
\end{align*}
Hence, $\bgr[m]$ can be expressed in terms of $\bp[m]$ and $\{\gammar[k], \taur[k]\}_{k\in \id{K}}$ as
\begin{equation}
\label{eq:convolution}
\bgr[m]= 
\bp[m] \bsr [m], \; \bsr[m] =\sum\limits_{k = 0}^{K - 1} \gammar\sqb{k}  \e^{-\jmath \frac{2m\pi \taur[k]}{\tau}}, \;\ m\is{Z}.
\end{equation}
Combining \eqref{eq:convolution} and \eqref{eq:bh}, we eventually derive the interplay between $\hgr$ and $\hp$ as
\begin{equation}
\label{eq:DFT model}
\hgr[l] = \tau \bgr [l] \eqr{eq:convolution} \hp[l] \bsr [l] = \hp[l] \sum\nolimits_{k = 0}^{K - 1} \gammar\sqb{k}  \e^{-\jmath \frac{2l\pi \taur[k]}{\tau}}, \; l\in\id{N}.
\end{equation}
\eqref{eq:DFT model} can be recast in time domain, which gives rise to a sparse representation
\begin{align}
\gr[n] &= \frac{1}{N}  \sum\nolimits_{l =0}^{N-1} \hgr [l] \zm{N}{n\cdot l} = \frac{1}{N}  \sum\nolimits_{l =0}^{N-1} \sum\nolimits_{u =0}^{N-1} \fr [u] \zm{N}{-u\cdot l} \bsr [l]  \zm{N}{n\cdot l} \notag \\
&=\sum\nolimits_{u =0}^{N-1} \fr [u] \rob{\frac{1}{N}  \sum\nolimits_{l =0}^{N-1} \bsr [l] \zm{N}{(n-u)\cdot l} }  =\sum\nolimits_{u =0}^{N-1} \fr [u]  \qr[\rob{n-u}_{{\rm mod}\;N}] 
\end{align}
where ${\rm mod}$ denotes modulo operation and $\qr$ is 
parametrized by $\{\taur[k], \gammar[k]\}_{k\in\id{K}}$ (see \fig{fig:Dirichlet Model}) as
\begin{align}
\qr[n] &= \frac{1}{N} \sum\nolimits_{l =0}^{N-1} \bsr [l] \zm{N}{n\cdot l} \EQc{eq:convolution} \frac{1}{N} \sum\nolimits_{k = 0}^{K - 1} \gammar\sqb{k} \sum\nolimits_{l =0}^{N-1}  e^{\jmath\frac{2l\pi }{N}(n - \frac{ \taur[k]}{T})}  \notag \\
&= \sum\limits_{k=0}^{K-1}  \frac{{{\Gamma _{\mathbf{r}}}\left[ k \right]} \big(1 - (u_k)^N \big)  }{ N(1 - u_k \zm{N}{n}) }  = \frac{\PMs (\zm{N}{n})}{\QMs (\zm{N}{n})}, \; \; 
\PMs \in P_{K-1}, \QMs \in P_{K}.
\end{align}
In the above, $\QMs$ is related to $\H$ in \eqref{eq:h filter} as
$
\QMs (z^{-1}) = \H(z) \Longleftrightarrow \QMs \rob{u_k^{-1}}=0,k\in\id{K}. 
$
\end{proof}

When considering the ToF kernels discussed in \secref{sec:SF}, the model in \eqref{eq:circ model} thus becomes an accurate \emph{approximation} due to the decaying Fourier spectrum of the kernels, as shown in \eqref{eq:decay}. This enables a flexible and practical framework that leverages the common kernel characteristics and recovers sparse spikes on a continuum.
More specifically, on the one hand, this derived model matches the various ToF modalities, which we validate via extensive hardware experiments in \secref{sec:exp}. On the other hand, this model leads to a tractable and efficient optimization method for the \srf and kernel recovery. 
Consequently, in the presence of distortions, \eg, quantization resolution, system noise and inter-reflections, the blind ToF imaging problem can be posed as
\begin{equation}
\label{eq:global problem}
\mathop {\min}
_{\substack{
\mphi,
\PMs,
\QMs 
}}
\     \norm{\mgr - \mphi \circov \mqr}_{2}^{2}, \; \mbox{s.t.} \;\; \qr[n] = \frac{\PMs (\zm{N}{n})}{\QMs (\zm{N}{n})},\;
\PMs \in P_{K-1},\; \QMs \in P_{K},\; n\in\id{N}.
\end{equation}
What makes the minimization in \eqref{eq:global problem} non-trivial is the structure and the constraints of the setup.  
In order to address this problem, we employ an alternating minimization strategy where the goal is to split \eqref{eq:global problem} into two tractable sub-problems:  $\PO$ that addresses the recovery of $\PMs,\QMs$ via non-linear model-fitting and $\PT$ that solves for $\mphi$ via least-squares fitting.

\subsection{Sub-Problem P1: Continuous-Time Spike Estimation}
\label{subsec:spike}
Assuming that $\mphi$ is known, it remains to estimate $\PMs,\QMs$ by solving the following quadratic minimization,
\begin{equation}
\label{eq:model fitting}
\PO \quad \quad  \mathop {\min}
_{\substack{
\PMs,\QMs
}}
\     \norm{\mgr - \mphi \circov \mqr}_{2}^{2},\; \mbox{s.t.} \;\; \qr[n] = \frac{\PMs (\zm{N}{n})}{\QMs (\zm{N}{n})},\;
\PMs \in P_{K-1},\; \QMs \in P_{K},\; n\in\id{N}.
\end{equation}
The bi-linearity of convolution operation results in
$$
\mphi \circov \mqr = \Toep{\fr} \mqr, \ [\Toep{\fr}]_{m,n} = \fr[(m-n)_{{\rm mod}\;N}].
$$
Then, $\PO$ in \eqref{eq:model fitting} translates to
\begin{equation}
\label{eq:ratio structure}
\mathop {\min}
_{\substack{
\PMs, \QMs
}}
\     \sum\limits_{m =0}^{N-1} \abs{\gr[m] - \sum\limits_{n=0}^{N-1}  \sqb{\Toep{\fr}}_{m,n} \tfrac{\PMs (\zm{N}{n})}{\QMs (\zm{N}{n})} }^{2}.
\end{equation}
As solving \eqref{eq:ratio structure} is challenging due to its non-linearity, we opt for an iterative strategy by 
constructing a collection of estimates for $\PMs,\QMs$, and selecting the one that minimizes the mean-squared error (MSE) of the measurements via \eqref{eq:ratio structure}. These estimates $\{\PMjj, \QMjj \}$ are found iteratively by solving the following approximate problem (since $\QMj\approx \QMs$)
\begin{align}
\label{eq:linear structure}
&\mathop {\min}
_{\substack{
\PMs, \QMs
}}
\  \sum\limits_{m =0}^{N-1} \abs{\ur[m]+ \sum\limits_{n=0}^{N-1} \sqb{\Toep{\fr}}_{m,n}  \tfrac{\qr^{[0]} [n] \QMs (\zm{N}{n}) -  \PMs (\zm{N}{n})}{\QMj (\zm{N}{n})}  }^{2} \notag \\
&\mbox{where} \ \ur[m] = \gr[m] - \sum\limits_{n=0}^{N-1}  \sqb{\Toep{\fr}}_{m,n} \qr^{[0]} [n] \notag \\
& \quad \quad \quad \  \ \mqr^{[0]} = \rob{\gram{\Toep{\fr}} }^{-1}  \Toep{\fr}^{\Hermit} \mgr, \quad
j\in \id{\jmax}
\end{align}
\rg{where $\mqr^{[0]}$ is an initial estimate of $\mqr$ and the resulting estimation error is characterized by the residue entry $\mur$. This structure ensures that the solution to \eqref{eq:linear structure} minimizes the model-fitting error defined in \eqref{eq:model fitting}.
Different initialization of ${\QMs^{[0]}}$ results to diverse estimates of ${\QMs^{[j]}}$, improving the accuracy and robustness of the algorithm. Here, we provide a deterministic initialization strategy: we pick the $K$ most prominent peaks of $\mqr^{[0]}$ \cite{Guo:2022:J}, which provides the roots of $\QMs^{[j]}$ as well as the polynomial coefficients. It is possible that the stopping criterion in \eqref{eq:stopping criterion} may not be met for this choice of ${\QMs^{[0]}}$ after reaching the maximum iteration count $\jmax$. In such cases, the algorithm is restarted with random initialized ${\QMs^{[0]}}$ that follows independent and identical unit Gaussian distributions. }

Having estimated $\QMs$ and $\PMs$, spike locations $\taur[k]$ are obtained by $\mathsf{roots}\rob{\QMs}\mapsto u_k^{-1}$ since $\QMs \rob{u_k^{-1}}=0$. The corresponding amplitudes $\gammar [k]$ are obtained via least-squares,
\begin{equation}
\label{eq:parameter estimates}
\taur[k]=-\tfrac{\tau}{2\pi} \mathsf{Im}\bigl( \log(u_k)\bigr)\quad \mbox{and} \quad \gammar[k] =-{\frac{{ N u_k {\PMs}\left( u_k^{-1} \right)}}{{\left( {1 - u_k^{N}} \right)
{\left. { \df{1}{\QMs}{z} \left( z \right)} \right|_{z = {u_k^{-1}}}}}}}.
\end{equation}
With $\{\taur[k], \gammar[k]\}_{k\in\id{K}}$, $\mqr$ can be reconstructed using \eqref{eq:circ model} and \ttd SRF can be recovered via \eqref{eq:SRF}.
\bpara{Algorithmic Implementation.} We provide an algorithm for \eqref{eq:linear structure}. The trigonometric polynomials $\PMs (\zm{N}{n}), \QMs (\zm{N}{n})$ in \eqref{eq:linear structure} can be algebraically written as, 
\[
\bigl[\PMs (\zm{N}{n})\bigr] = \vmat{N}{K} \p
\ \mbox{ and } \
\bigl[\QMs(\zm{N}{n})\bigr] = \vmat{N}{K+1} \q
\]
where $\p$ and $\q$ are the coefficients of $\PMs \in P_{K-1}$ and $\QMs \in P_{K}$. Assuming the estimate $\q^{[j]}$ of $\q$ at iteration-$j$ is known, the minimization at iteration-$j+1$ can be reformulated in matrix form as,  
\begin{align}
\label{eq:linearization}
\{\p^{[j+1]},\q^{[j+1]}\}=\mathop{\rm arg\,min}\limits_{\p, \q}
\  \norm{\mur + \Aj \q - \Bj \p}_{2}^{2}  
\end{align}
where the entities $\Aj$, $\Bj$ and $\mur$ are respectively given by,
\begin{equation}
\label{eq:matrices}
\begin{array}{*{20}{l}}
\Aj= \Toep{\fr} \Rj \dk (\mqr^{[0]}) \vmat{N}{K+1}
&
\Bj=  \Toep{\fr} \Rj \vmat{N}{K}, \quad 
\mur = \mgr - \Toep{\fr} \rob{\gram{\Toep{\fr}} }^{-1}  \Toep{\fr}^{\Hermit} \mgr \\ 
\hfill \mbox{ and in the above, the variables }
& 
\Rj = \rob{\dk(\vmat{N}{K+1} \q^{[j]})}^{-1}, \; 
\mbox{and} \;\; \mqr^{[0]} = \rob{\gram{\Toep{\fr}} }^{-1}  \Toep{\fr}^{\Hermit} \mgr.
\end{array}
\end{equation}
We adopt a normalization constraint below to ensure the uniqueness of the optimal solution to \eqref{eq:linearization}  
\begin{equation}
\label{eq:normalization}
\frac{1}{2\pi} \int_{0}^{2\pi}\rob{\QMs^{[0]}(e^{-\jmath\theta})}^{\ast}\QMs^{[j+1]}(e^{-\jmath\theta})d\theta=1 
\end{equation}
where $(\cdot)^{\ast}$ denotes conjugation operator and $\QMs^{[0]} $ is the initialization of the algorithm. 
Consequently, the quadratic minimization~\eqref{eq:linearization} can be posed as
\begin{align}
&\{\p^{[j+1]},\q^{[j+1]}\}=\mathop{\rm arg\,min}\limits_{\p, \q}
\  \norm{\mur + \Aj \q - \Bj \p}_{2}^{2}  \\
&\mbox{subject to}  \quad  \frac{1}{2\pi} \int_{0}^{2\pi}\rob{\QMs^{[0]}(e^{-\jmath\theta})}^{\ast}\QMs^{[j+1]}(e^{-\jmath\theta})d\theta=1 \notag
\end{align}
for which its optimal solution can be found by solving the following system of linear equations
\begin{equation}
\label{eq:optimal solution}
\begin{bmatrix}
\gram{\Cj}  &\xo \\
(\xo)^{\Hermit}&0
\end{bmatrix} 
\begin{bmatrix}
\xjj\\
\lambda
\end{bmatrix} 
= 
\begin{bmatrix}
(\Cj)^{\Hermit}\mur \\
1
\end{bmatrix}\;
\mbox{and} 
\begin{array}{*{20}{l}}
& \xjj = 
\begin{bmatrix}
\q^{[j+1]} \\
\p^{[j+1]}
\end{bmatrix},
\; \xo = 
\begin{bmatrix}
\q^{[0]} \\
\mathbf{0}
\end{bmatrix} , \\ 
& \Cj=[-\Aj,\;\Bj]
\end{array}
\end{equation}
where $\q^{[0]}$ are the coefficients of $\QMs^{[0]} \in P_{K}$ and $\lambda$ is the Lagrange multiplier such that the normalization constraint~\eqref{eq:normalization} is satisfied\footnote{For each initialization $\q^{[0]}$, we update the trigonometric polynomial coefficients $\q^{[j+1]}$ until the stopping criterion~\eqref{eq:stopping criterion} is met. If \eqref{eq:stopping criterion} is not met for certain choices of $\q^{[j+1]}$ after reaching the maximum iteration count $\jmax$, we restart the algorithm with a different initialization.}.

\begin{algorithm}[!t]
\caption{Blind ToF Imaging Algorithm}
\label{alg:sparse recovery}
\begin{algorithmic}[1]
\REQUIRE Noisy ToF measurements $\mgr$.
\STATE   \noindent \textbf{Kerne Initialization}: Compute $\mphi^{[0]}$ via \eqref{eq:ini}. 
\FOR{\texttt{loop = 1} to \texttt{max. iterations}}
\FOR{${j = 1}$ to $\jmax$}
\STATE Construct the matrices in~\eqref{eq:matrices}; 
\STATE Update $\p^{[j]},\q^{[j]}$ by solving~\eqref{eq:optimal solution};
\IF{\eqref{eq:stopping criterion} holds}
\STATE \texttt{Terminate all loops};
\ENDIF
\ENDFOR 
\STATE Calculate $\{\taur[k], \gammar[k]\}_{k\in\id{K}}$ using~\eqref{eq:parameter estimates}; 
\STATE Update $\mphi$ using \eqref{eq:opt phi};
\IF{\eqref{eq:stopping criterion} holds}
\STATE \texttt{Terminate all loops};
\ENDIF
\ENDFOR
\ENSURE The \srf parameters $\{\taur[k], \gammar[k]\}_{k\in\id{K}}$ and the kernel $\mphi$. 
\end{algorithmic}
\end{algorithm}

\subsection{Sub-Problem P2: Kernel Recovery} 
\label{subsec:kernel}
With $\mqr$ known from the method in \secref{subsec:spike}, the minimization on $\mphi$ essentially boils down to the least-squares problem, resulting in the solution
\begin{equation}
\label{eq:opt phi}
\mphi = \rob{\gram{\Toep{\mqr}} }^{-1}  \Toep{\mqr}^{\Hermit} \mgr.
\end{equation}

\bpara{Stopping Criterion.} We initialize the proposed method in \algref{alg:sparse recovery} by computing,
\begin{equation}
\label{eq:ini}
\mphi^{[0]} = \mathsf{Re}\rob{\vmat{N}{N} (\dk (e^{\jmath \bf w}) \wmat{N}{N} \mgr )}, \qquad \rgnn{\mat{w} \sim \mathcal{N}\rob{{\bf 0},\I}}
\end{equation}
which is a reasonable initialization based on kernel features. Using \eqref{eq:linearization}, we then estimate $\{\taur[k], \gammar[k]\}_{k\in\id{K}}$ via \eqref{eq:parameter estimates} based on which we refine $\mphi$ via \eqref{eq:opt phi}. We use the following
\begin{equation}
\label{eq:stopping criterion}
\norm{\mgr - \mphi \circov \mqr}_{2}  \leqslant \sigma
\end{equation}
as our stopping criterion, where $\sigma$ represents the data distortion level. In other words, we can only recover the signal up to a tolerance level of $\sigma$.
Iterating the method offers robust estimates with super-resolution capability, which is validated via hardware experiments.
\rgn{
Empirically, $10$ random initializations and $\jmax=20$ are sufficient to obtain an accurate solution that satisfies the stopping criterion in \eqref{eq:stopping criterion}, provided that the spikes $\{\taur[k] \}_{k\in\id{K}}$ are relatively well-separated. In the challenging scenarios of resolving closely-located spikes, more restarts with different initializations ($\approx 20$) are required to find a reasonable solution that fits the ToF data within a certain tolerance level $\sigma$. We would like to point out that the algorithmic complexity can be potentially improved by 
\begin{enumerate*}[leftmargin = *, label = \uline{\arabic*})]
\item leveraging algebraic structures of matrices $ \Toep{\fr}, \Toep{\mqr}$ and shared features of $\mphi$
\item optimizing numerical calculation and implementation.
\item utilizing effective dimensionality reduction techniques.
\end{enumerate*}
We tabulate the algorithm run-time in the last column of \tabref{tab:exp}.} 
The algorithm is summarized in \algref{alg:sparse recovery}.

\rgn{While our method assumes the model order ($K$ in \eqref{eq:filtered samples}) is known, this is not strictly necessary. A practical approach leverages the underlying physics of the imaging process. Since light reflections decay according to the inverse-square law, for $K > 3$, the reflections are likely submerged in the noise floor. Thus, the algorithm can be implemented with $K = 4$, where any spurious spikes from overestimating the model order will reflect system noise. These will correspond to relatively smaller coefficients $\Gamma_{\mathbf{r}}$ in \eqref{eq:filtered samples} and can be removed through hard-thresholding. Alternatively, $\sigma$ in \eqref{eq:stopping criterion} measures the quality of fitting. Hence, one may develop a method that progressively fits the data until the stopping criterion \eqref{eq:stopping criterion} is met. Lastly, algorithms such as SORTE\footnote{stands for Second ORder sTatistic of the Eigenvalues.} \cite{He:2010:J} can also be combined with our approach to incorporate model-order estimation.}

\begin{table*}[!t]
\centering
\caption{Hardware Based Experimental Parameters and Performance Evaluation.}
\label{tab:exp}
\resizebox{\columnwidth}{!}{%
\begin{tabular}{@{}lcccccccccccc@{}}
\toprule
Figure & Exp. No. & $\mathbf{r}$ & $K$ & $T$&${\Gamma _{\mathbf{r}}}\left[ k \right]$       & ${\tau _{\mathbf{r}}}\left[ k \right]$& ${\widetilde \Gamma _{\mathbf{r}}}\left[ k \right]$ & ${\widetilde \tau _{\mathbf{r}}}\left[ k \right]$ & \multicolumn{2}{c}{MSE} &$\PSNR{\bm\fr}{\bm\varphi}$ &Run-Time  \\ \midrule
&  &  &   & ($\times 10^{-12}$ sec.) &     &  ($\times 10^{-8}$ sec.) &   & ($\times 10^{-8}$ sec.)  &  & &(dB) & (s)\\ \midrule
&               &               &                       &                       & \multicolumn{2}{c}{\cellcolor[HTML]{f3ff9c}Known kernel} & \multicolumn{2}{c}{\cellcolor[HTML]{ffec9c}Blind} &    $\msem{\Gamma}{\Gamma}$           &        $\msem{\bm\tau}{\bm\tau}$      \\ \midrule
\fig{fig:ToFkernel}&  \rom{1}                    &    $[60\;60]^{\top}$                  &      2     &$70$            &         [1.19,0.23]                    &         [8.45,9.44]                   &    [1.19,0.23]                     &        [8.44,9.44]                 &   $2.31 \times10^{-5}$            &    $1.87 \times10^{-4}$  &      43.24   & \rgn{35.88}   \\
\fig{fig:movingSR} (a)&   \rom{2}-(a)          & $[60\;60]^{\top}$ & 2       &$96.15$              &      [0.86,0.52]                &      [3.03,4.19]                       &  [0.86,0.53]                          &    [3.02,4.19]                    &     $3.57 \times10^{-5}$                 &       $1.16 \times 10^{-4}$   &    41.93  & \rgn{37.07}   \\
\fig{fig:movingSR} (b)&    \rom{2}-(b)        & $[60\;60]^{\top}$  & 2      &$96.15$               &      [0.64,0.54]                 &      [3.23,4.20]                       &  [0.66,0.53]                          &    [3.21,4.20]                     &     $1.69 \times10^{-4}$                 &       $1.40 \times 10^{-4}$   &    40.29  & \rgn{36.63}   \\
\fig{fig:movingSR} (c)&    \rom{2}-(c)     & $[60\;60]^{\top}$                      &   2       &$96.15$             &  [0.48,0.59]                           &   [3.42,4.19]                         &  [0.58,0.46]                       &  [3.43,4.18]                      &     $1.35 \times 10^{-2}$           &  $7.13 \times 10^{-5}$ &       39.71    & \rgn{117.54}     \\
\fig{fig:movingSR} (d)&    \rom{2}-(d)     & $[60\;60]^{\top}$                     &       2    &$96.15$            &       [0.34,0.58]                      &    [3.64,4.20]                        &     [0.48,0.43]                    &          [3.62,4.20]              &    $2.13 \times 10^{-2}$            &     $1.90 \times 10^{-4}$  &     39.18  & \rgn{122.32}    \\
\fig{fig:3spikes} (b)&    \rom{3}-(a)      &  $[50\;70]^{\top}$                     &      3     &$70$            &    [0.70,0.44,0.13]                       &       [8.44,9.65,10.95]                  &      [0.66,0.48,0.16]                  &   [8.44,9.65,10.95]                      &      $1.50 \times 10^{-3}$         &    $1.57 \times 10^{-5}$ &      36.33   & \rgn{114.97}       \\
\fig{fig:3spikes} (c)&    \rom{3}-(b)                  &   $[45\;60]^{\top}$                    &         3     &$70$         &  [0.29,0.36,0.21]                           &     [8.51,9.71,11.08]                    &   [0.28,0.36,0.23]                    &    [8.53,9.74,11.08]                    &     $1.52 \times 10^{-4}$             &     $4.70 \times 10^{-4}$      &     36.22 & \rgn{106.66}   \\
\fig{fig:spad} (b)&     \rom{4}-(a)                    &  \textemdash                     &       2      &$6.1$          &     [1.34,1.72]                        &   [1.11,1.24]                         &      [1.33,1.72]                   &   [1.11,1.24]                      &      $2.61 \times 10^{-5}$            &    $2.33 \times 10^{-8}$ &      47.72  & \rgn{54.54}   \\
\fig{fig:spad} (c)&      \rom{4}-(b)                   &  \textemdash                     &        2     &$6.1$          &    [1.39,1.29]                         &    [1.18,1.24]                        &   [1.37,1.31]                      & [1.18,1.24]                       &          $3.26 \times 10^{-4}$        &      $7.19 \times 10^{-7}$  &      43.64 & \rgn{171.99} \\
\fig{fig:spad} (d)&      \rom{4}-(c)           &  \textemdash                     &        2    &$6.1$           & [1.69,0.89]                  &  [1.22,1.24]                         &    [1.83,0.54]           &  [1.22,1.24]          &      $ 7.09 \times 10^{-2}$            &      $9.31 \times 10^{-7}$    &      42.24  & \rgn{249.01}
\\ \bottomrule
\end{tabular}%
}
\end{table*}

\begin{figure}[!tb]
\centering
\scalebox{0.8}{
  
\tikzset {_ippmrayr8/.code = {\pgfsetadditionalshadetransform{ \pgftransformshift{\pgfpoint{0 bp } { 0 bp }  }  \pgftransformscale{1 }  }}}
\pgfdeclareradialshading{_a4d08zhd0}{\pgfpoint{0bp}{0bp}}{rgb(0bp)=(1,1,1);
rgb(16.25bp)=(1,1,1);
rgb(25bp)=(0.9,0.9,0.9);
rgb(400bp)=(0.9,0.9,0.9)}

  
\tikzset {_ua0vq8tpx/.code = {\pgfsetadditionalshadetransform{ \pgftransformshift{\pgfpoint{0 bp } { 0 bp }  }  \pgftransformscale{1 }  }}}
\pgfdeclareradialshading{_k06ii8nll}{\pgfpoint{0bp}{0bp}}{rgb(0bp)=(1,1,1);
rgb(16.25bp)=(1,1,1);
rgb(25bp)=(0.9,0.9,0.9);
rgb(400bp)=(0.9,0.9,0.9)}
\tikzset{every picture/.style={line width=0.75pt}} 

\begin{tikzpicture}[x=0.75pt,y=0.75pt,yscale=-1,xscale=1]

\path  [shading=_a4d08zhd0,_ippmrayr8]  (237.5, 39) circle [x radius= 87.68, y radius= 26.87]   ; 
 \draw  [color={rgb, 255:red, 223; green, 25; blue, 21 }  ,draw opacity=1 ][line width=1.5]   (237.5, 39) circle [x radius= 87.68, y radius= 26.87]   ; 

\draw (237.5,39) node  [font=\small] [align=left] {\begin{minipage}[lt]{84.32pt}\setlength\topsep{0pt}
\begin{center}
\textbf{{\fontfamily{\sfdefault}\selectfont Multi-path Imaging}}\\ \secref{subsec:Multi-path Imaging} 
\end{center}

\end{minipage}};
\path  [shading=_k06ii8nll,_ua0vq8tpx]  (471, 39) circle [x radius= 111.72, y radius= 26.87]   ; 
 \draw  [color={rgb, 255:red, 223; green, 25; blue, 21 }  ,draw opacity=1 ][line width=1.5]   (471, 39) circle [x radius= 111.72, y radius= 26.87]   ; 

\draw (471,39) node  [font=\small] [align=left] {\begin{minipage}[lt]{107.27pt}\setlength\topsep{0pt}
\begin{center}
\textbf{{\fontfamily{\sfdefault}\selectfont  Light-in-Flight Imaging }}\\ \secref{subsec:Light-in-Flight Imaging} 
\end{center}

\end{minipage}};
\draw  [draw opacity=0][fill={rgb, 255:red, 74; green, 144; blue, 226 }  ,fill opacity=0.2 ][line width=0.75]   (47.5,124) -- (145.5,124) -- (145.5,162) -- (47.5,162) -- cycle  ;
\draw (96.5,143) node  [font=\footnotesize,color={rgb, 255:red, 0; green, 21; blue, 255 }  ,opacity=1 ] [align=left] {\begin{minipage}[lt]{63.77pt}\setlength\topsep{0pt}
\begin{center}
{\fontfamily{\sfdefault}\selectfont  Diffuse Imaging }\\ \secref{subsubsec:Diffuse Imaging}
\end{center}

\end{minipage}};
\draw  [draw opacity=0][fill={rgb, 255:red, 74; green, 144; blue, 226 }  ,fill opacity=0.2 ][line width=0.75]   (169.5,124) -- (305.5,124) -- (305.5,162) -- (169.5,162) -- cycle  ;
\draw (237.5,143) node  [font=\footnotesize,color={rgb, 255:red, 0; green, 21; blue, 255 }  ,opacity=1 ] [align=left] {\begin{minipage}[lt]{100pt}\setlength\topsep{0pt}
\begin{center}
{\fontfamily{\sfdefault}\selectfont  Super-resolution (K=2) }\\  \secref{subsubsec:K=2}
\end{center}

\end{minipage}};
\draw  [draw opacity=0][fill={rgb, 255:red, 74; green, 144; blue, 226 }  ,fill opacity=0.2 ][line width=0.75]   (325,124) -- (485,124) -- (485,162) -- (325,162) -- cycle  ;
\draw (405,143) node  [font=\footnotesize,color={rgb, 255:red, 0; green, 21; blue, 255 }  ,opacity=1 ] [align=left] {\begin{minipage}[lt]{130pt}\setlength\topsep{0pt}
\begin{center}
{\fontfamily{\sfdefault}\selectfont  Higher order Imaging (K>2) }\\ \secref{subsubsec:K=3}
\end{center}

\end{minipage}};
\draw  [color={rgb, 255:red, 245; green, 166; blue, 35 }  ,draw opacity=1 ][fill={rgb, 255:red, 248; green, 231; blue, 28 }  ,fill opacity=0.8 ][line width=1.5]   (118,218) .. controls (118,212.48) and (122.48,208) .. (128,208) -- (212,208) .. controls (217.52,208) and (222,212.48) .. (222,218) -- (222,222) .. controls (222,227.52) and (217.52,232) .. (212,232) -- (128,232) .. controls (122.48,232) and (118,227.52) .. (118,222) -- cycle  ;
\draw (170,220) node [font=\footnotesize] [align=left] 
{\begin{minipage}[lt]{80pt}\setlength\topsep{0pt}
\begin{center}
{Lock-in Sensors \cite{Foix:2011:J}}
\end{center}

\end{minipage}};
\draw  [color={rgb, 255:red, 245; green, 166; blue, 35 }  ,draw opacity=1 ][fill={rgb, 255:red, 248; green, 231; blue, 28 }  ,fill opacity=0.8 ][line width=1.5]   (258,218) .. controls (258,212.48) and (262.48,208) .. (268,208) -- (322,208) .. controls (327.52,208) and (332,212.48) .. (332,218) -- (332,222) .. controls (332,227.52) and (327.52,232) .. (322,232) -- (268,232) .. controls (262.48,232) and (258,227.52) .. (258,222) -- cycle  ;
\draw (295,220) node  [font=\footnotesize] [align=left] {\begin{minipage}[lt]{47.6pt}\setlength\topsep{0pt}
\begin{center}
{TSCPC \cite{Pellegrini:2000:J}}
\end{center}

\end{minipage}};
\draw    (181.68,59.72) .. controls (140.85,68.77) and (115.41,89.7) .. (105.38,122.49) ;
\draw [shift={(104.93,124)}, rotate = 286.09] [color={rgb, 255:red, 0; green, 0; blue, 0 }  ][line width=0.75]    (10.93,-3.29) .. controls (6.95,-1.4) and (3.31,-0.3) .. (0,0) .. controls (3.31,0.3) and (6.95,1.4) .. (10.93,3.29)   ;
\draw    (237.5,65.87) -- (237.5,122) ;
\draw [shift={(237.5,124)}, rotate = 270] [color={rgb, 255:red, 0; green, 0; blue, 0 }  ][line width=0.75]    (10.93,-3.29) .. controls (6.95,-1.4) and (3.31,-0.3) .. (0,0) .. controls (3.31,0.3) and (6.95,1.4) .. (10.93,3.29)   ;
\draw    (303.26,56.77) .. controls (343.68,59.9) and (373.68,81.89) .. (393.25,122.76) ;
\draw [shift={(393.84,124)}, rotate = 244.85] [color={rgb, 255:red, 0; green, 0; blue, 0 }  ][line width=0.75]    (10.93,-3.29) .. controls (6.95,-1.4) and (3.31,-0.3) .. (0,0) .. controls (3.31,0.3) and (6.95,1.4) .. (10.93,3.29)   ;
\draw    (333.18,37.5) -- (351.28,37.5)(333.18,40.5) -- (351.28,40.5) ;
\draw [shift={(359.28,39)}, rotate = 180] [color={rgb, 255:red, 0; green, 0; blue, 0 }  ][line width=0.75]    (10.93,-3.29) .. controls (6.95,-1.4) and (3.31,-0.3) .. (0,0) .. controls (3.31,0.3) and (6.95,1.4) .. (10.93,3.29)   ;
\draw [shift={(325.18,39)}, rotate = 360] [color={rgb, 255:red, 0; green, 0; blue, 0 }  ][line width=0.75]    (10.93,-3.29) .. controls (6.95,-1.4) and (3.31,-0.3) .. (0,0) .. controls (3.31,0.3) and (6.95,1.4) .. (10.93,3.29)   ;
\draw  [dash pattern={on 4.5pt off 4.5pt}]  (448.81,65.34) .. controls (401.73,144.09) and (318.34,31.64) .. (253.94,122.61) ;
\draw [shift={(252.97,124)}, rotate = 304.61] [color={rgb, 255:red, 0; green, 0; blue, 0 }  ][line width=0.75]    (10.93,-3.29) .. controls (6.95,-1.4) and (3.31,-0.3) .. (0,0) .. controls (3.31,0.3) and (6.95,1.4) .. (10.93,3.29)   ;
\draw    (202.13,162) .. controls (183.42,165.6) and (173.3,180.39) .. (171.75,206.39) ;
\draw [shift={(171.67,208)}, rotate = 272.59] [color={rgb, 255:red, 0; green, 0; blue, 0 }  ][line width=0.75]    (10.93,-3.29) .. controls (6.95,-1.4) and (3.31,-0.3) .. (0,0) .. controls (3.31,0.3) and (6.95,1.4) .. (10.93,3.29)   ;
\draw    (267.49,162) .. controls (285.7,167.73) and (294.63,182.45) .. (294.25,206.16) ;
\draw [shift={(294.21,208)}, rotate = 272.02] [color={rgb, 255:red, 0; green, 0; blue, 0 }  ][line width=0.75]    (10.93,-3.29) .. controls (6.95,-1.4) and (3.31,-0.3) .. (0,0) .. controls (3.31,0.3) and (6.95,1.4) .. (10.93,3.29)   ;

\end{tikzpicture}}  

\caption{Roadmap for hardware-based ToF imaging experiments. We demonstrate the performance of the proposed blind ToF imaging approach across various experimental setups and data modalities.
}
\label{fig:exp map}
\end{figure}
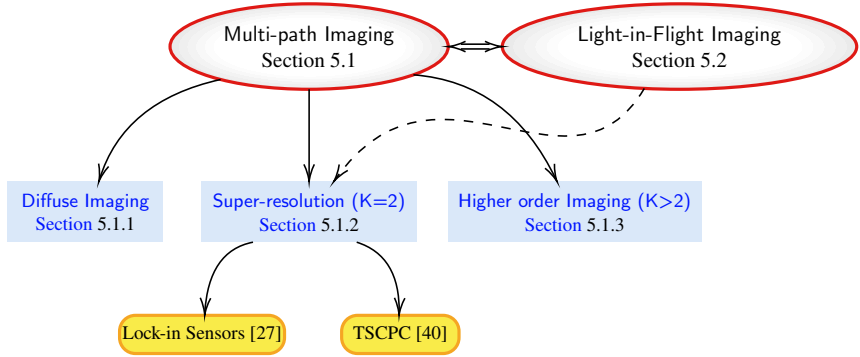

\section{Experiments}
\label{sec:exp}
Our experiments aim to validate the effectiveness of the blind ToF imaging approach, demonstrating that it performs comparably to methods with prior kernel calibration, which we use as a baseline for ground truth. Through a series of $10$ experiments, we achieve robust 3D scene reconstruction across various experimental setups and ToF datasets without any kernel calibration. \fig{fig:exp map} illustrates the different ToF imaging scenarios included in our experiments. We assess performance using the following quantitative metrics. Mean-Squared Error or MSE $\msem{\Gamma}{\Gamma}$ and $\msem{\bm\tau}{\bm\tau}$ is used to evaluate the \ttd scene reconstruction accuracy. Peak signal-to-noise ratio (PSNR) is utilized to assess the accuracy of kernel estimation, \ie,
$
\PSNR{\bm \fr}{\bm \varphi} = 10\log_{10} \rob{\frac{\norm{\fr }_{\infty}^{2} }{\msep{\bm\fr}{\bm\varphi} }}.
$
We tabulate the experimental parameters and performance evaluation in \tabref{tab:exp}. 

\subsection{Multi-path Imaging}
\label{subsec:Multi-path Imaging}
\begin{figure}
\begin{center}
\includegraphics[width=0.55\textwidth]{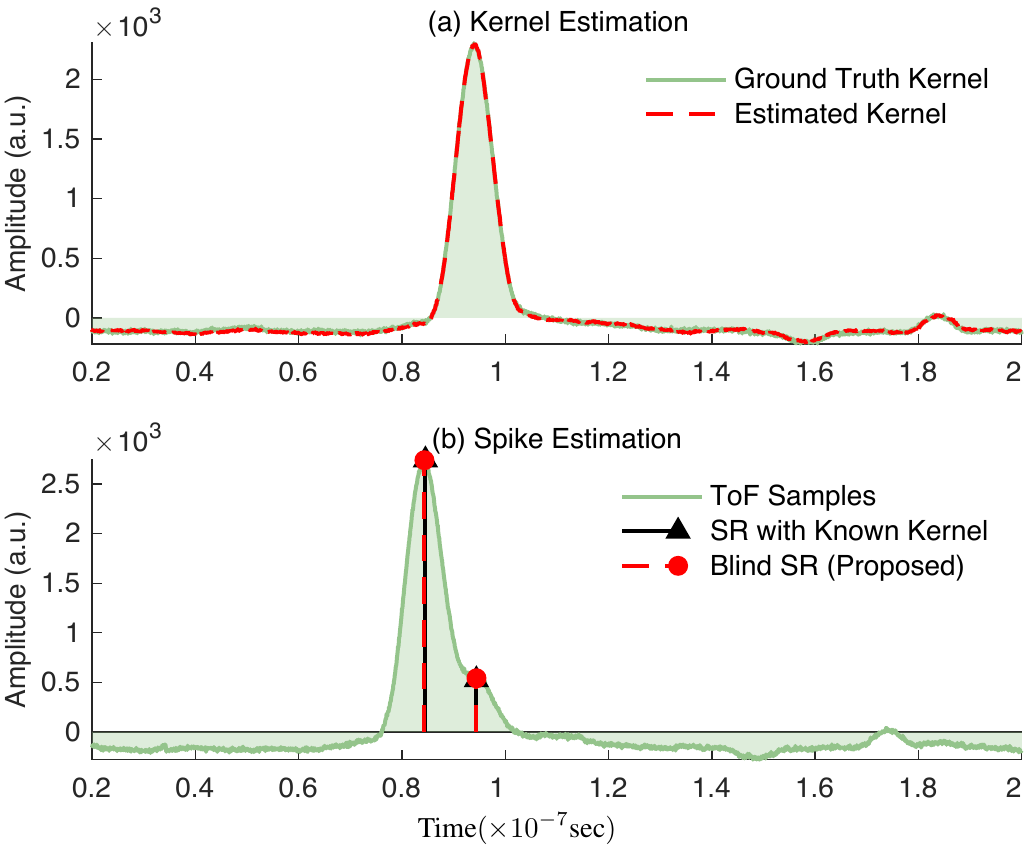} 	
\end{center}
\caption{Blind recovery of single pixel ToF measurements ($\mathbf{r} = \left[60\;60\right]^{\top}$, $K=2$). (a) Kernel reconstruction. (b) \ttd \srf estimation. The experimental setup is shown in \fig{fig:LIF}.
}
\label{fig:ToFkernel}  
\end{figure}
Commercial ToF systems determine distance by amplitude modulation of a continuous wave. Although ToF sensors typically provide a single depth value per pixel, the actual scene might contain multiple depths, such as a semi-transparent surface in front of a wall. This situation is known as a mixed pixel scenario \cite{Kadambi:2013:J, Bhandari:2016:J}, where multiple light paths converge at the same pixel, resulting in a measured range that is a nonlinear mixture of the incoming paths.

To address the issue of multi-path interference, one effective strategy involves emitting a custom code and capturing a series of demodulated values through successive electronic delays. Subsequently, a sparse deconvolution \cite{Kadambi:2013:J, Bhandari:2016:J} is applied to isolate a sequence of Diracs in the time profile, each corresponding to different light path depths and multi-path combinations. We consider the results from this method as the ground truth for validating our blind recovery approach on these ToF datasets.

\subsubsection{Diffuse Imaging}
\label{subsubsec:Diffuse Imaging}

We use the setup from \cite{Kadambi:2013:J, Bhandari:2016:J}: The scene consists of a placard reading ``TIME OF FLIGHT'' which is hidden by a diffusive semi-translucent sheet. Raw data comprising of a $120\times 120\times 2976$ image tensor is acquired. Here, $N=2976$ refers to the equidistant/uniform ToF measurements captured with sampling time $T=70$ ps. The ToF measurements have been acquired with a custom ToF camera equipped with a \texttt{PMD 19k-S3 sensor} utilizing the architecture previously used in \cite{Kadambi:2013:J,Bhandari:2015:J,Bhandari:2016:J}. Both the illumination and the ToF pixels are modulated using the same binary M-sequences. This leads to a relatively narrow cross-correlation function in time domain. The illumination control signal can be shifted with regard to the pixel control signal, thus shifting the resulting cross-correlation function. 
\begin{figure}
\begin{center}
\begin{overpic}[width=0.55\columnwidth]{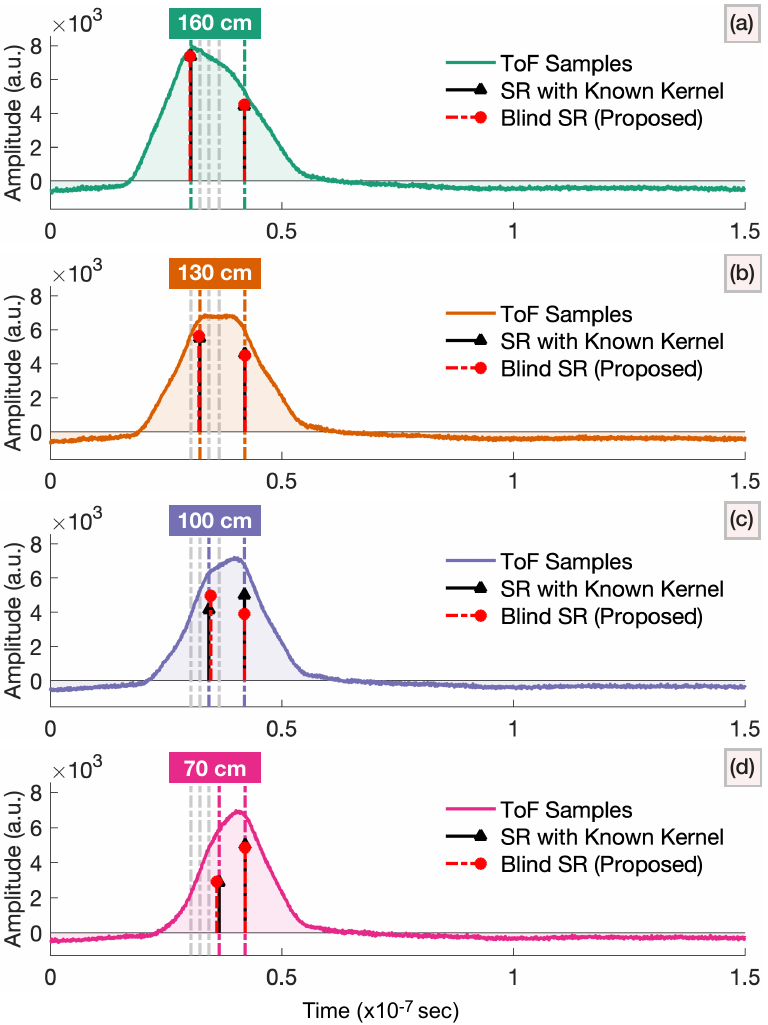} 	
\put (30,97) {\colorbox{white}{{\fontsize{9pt}{9pt}\selectfont {\color{black} \href{https://youtu.be/wMlWJv7B66o}{\texttt{https://youtu.be/wMlWJv7B66o}} }}}}
\put (30,48.5) {\colorbox{white}{{\fontsize{9pt}{9pt}\selectfont {\color{black} \href{https://youtu.be/F-g6X85DWO4}{\texttt{https://youtu.be/F-g6X85DWO4}} }}}}
\put (30,24) {\colorbox{white}{{\fontsize{9pt}{9pt}\selectfont {\color{black} \href{https://youtu.be/fO_4ivWC2Hg}{\texttt{https://youtu.be/fO\_4ivWC2Hg}} }}}}
\end{overpic}
\end{center}
\caption{
Lock-in sensor based super-resolved ToF imaging. 
We benchmark the performance of the proposed method with kernel calibration. 
The scene consists of a mannequin head placed between a diffusive semi-translucent surface and a wall in the backdrop. By moving the diffusive surface, the inter-object separation is uniformly reducing from (a) $160$, (b) $130$, (c) $100$ to (d) $70$ cm, respectively. The dashed lines represent the ground truth spikes. 
Our method super-resolves the inter-object separation for (a) $161$, (b) $134$, (c) $98$ and (d) $72$ cm, which accurately matches the experimental setup in \fig{fig:LIF}.}
\label{fig:movingSR} 
\end{figure}
To contextualize our results, we first show the blind recovery of a single-pixel ToF data ($\mathbf{r} = \left[60\;60\right]^{\top}$) in \fig{fig:ToFkernel}. The estimated kernel and reconstructed pixel measurements reach accuracy with $\PSNR{\bm\fr}{\bm\varphi} = 43.24$ dB, $\msem{\Gamma}{\Gamma} = 2.31 \times10^{-5}$, and $\msem{\bm\tau}{\bm\tau}=1.87 \times10^{-4}$. Then, we reconstruct the \ttd scene from estimated time profiles by running the blind recovery algorithm for each pixel measurements. This leads to indistinguishable results compared to the ground truth shown in \fig{fig:3DToF}.

\begin{figure}[!t]
\centering
\includegraphics[width=1\columnwidth]{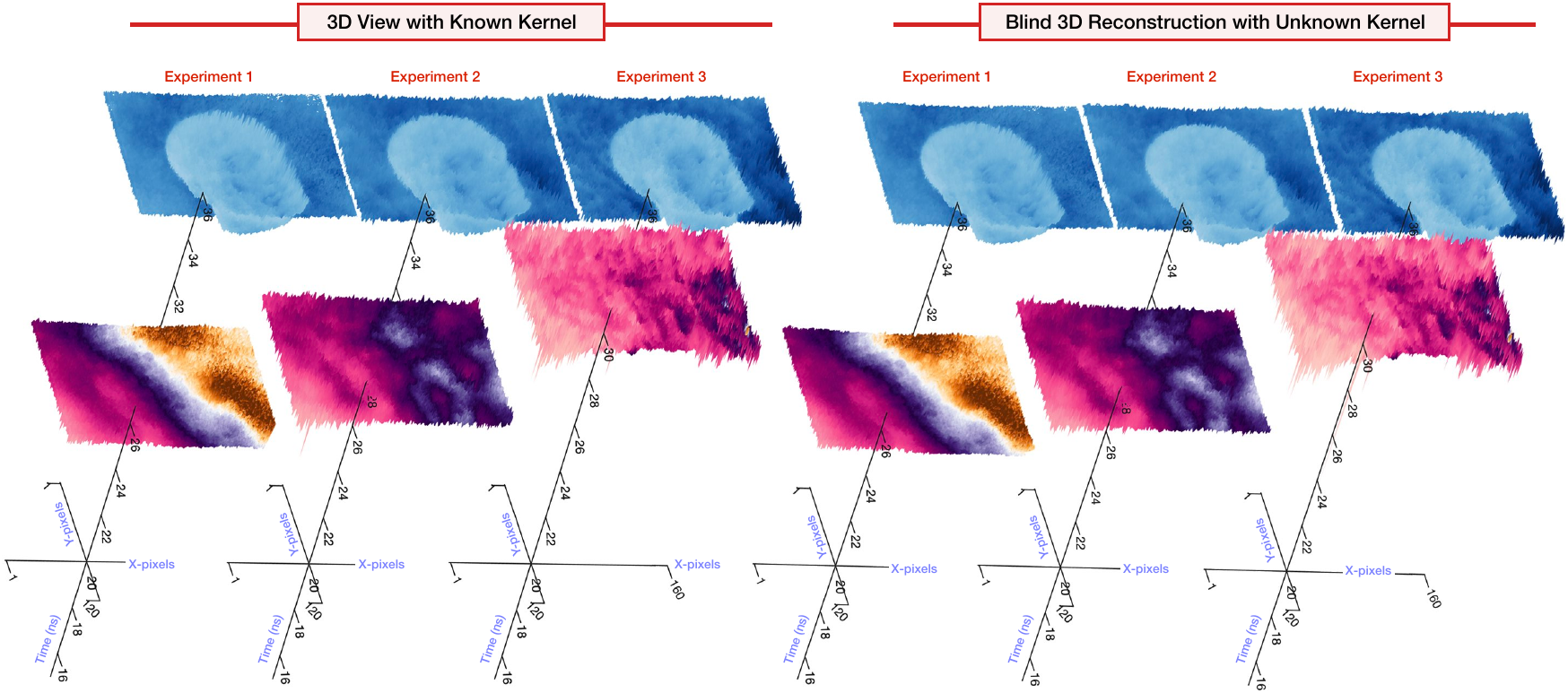} 		
\caption{Benchmarking blind ToF imaging with kernel calibration. The distance between the translucent surface and the mannequin head is reducing from $160$, $100$ to $70$ cm, respectively, as shown in \fig{fig:LIF}. This requires algorithmic \SR capability.
(a) \ttd visualization of depth imaging with kernel calibration. (b) \ttd visualization of blind depth imaging using the proposed blind recovery method.}
\label{fig:headSR}
\end{figure}

\begin{figure}[!t]
\centering
\includegraphics[width=0.8\columnwidth]{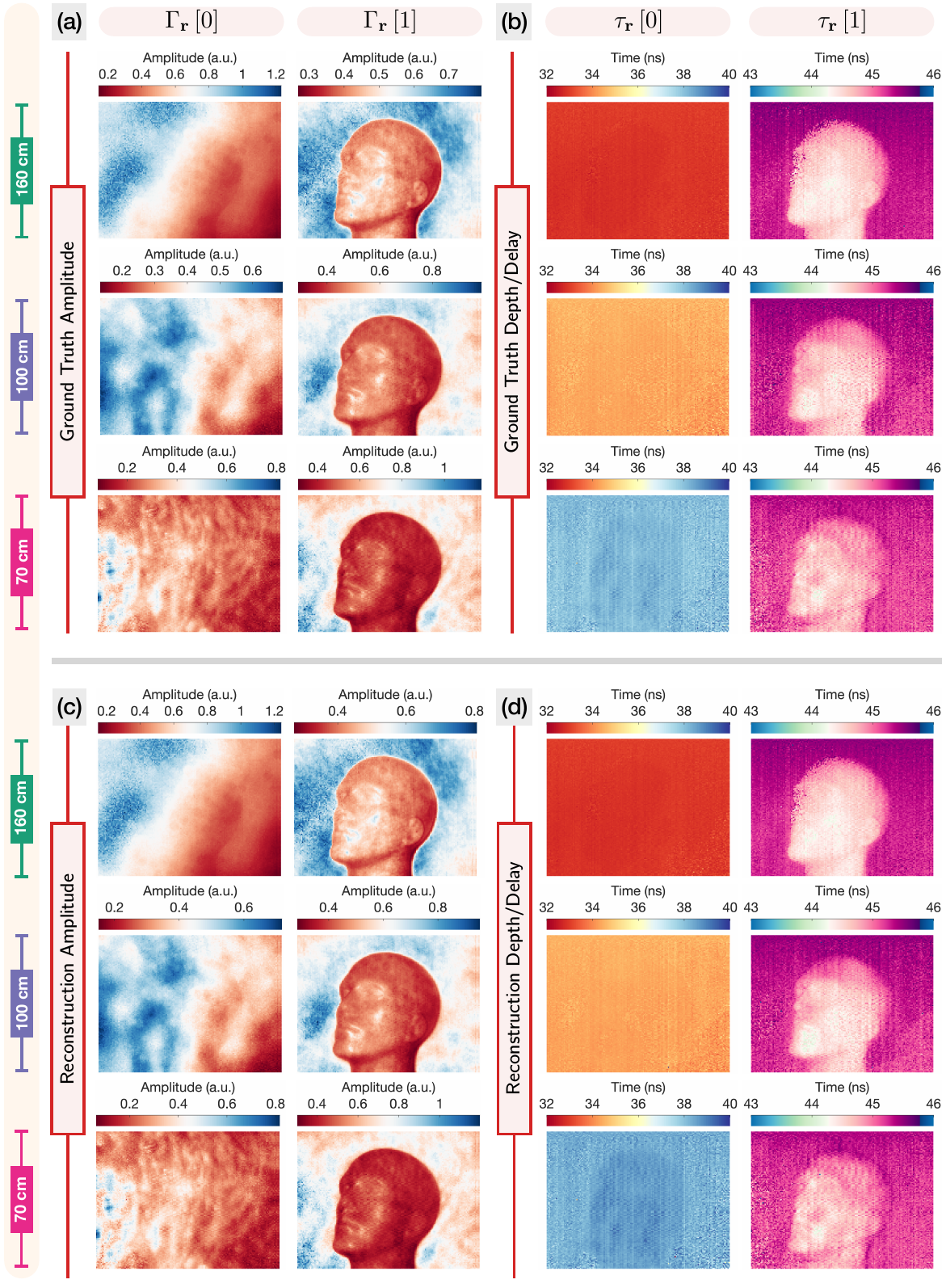} 		
\caption{Super-resolving object through a diffusive semi-translucent sheet. (a) and (b) are amplitude and depth imaging using known kernel; (c) and (d) are corresponding results utilizing our blind recovery method. The experimental setup is shown in \fig{fig:LIF}. Despite the challenges of small inter-object separation, our method still achieves accurate scene reconstructions in all scenarios.
}
\label{fig:ToFSR}
\end{figure}

\begin{figure}[!t]
\begin{center}
\includegraphics[width=0.55\columnwidth]{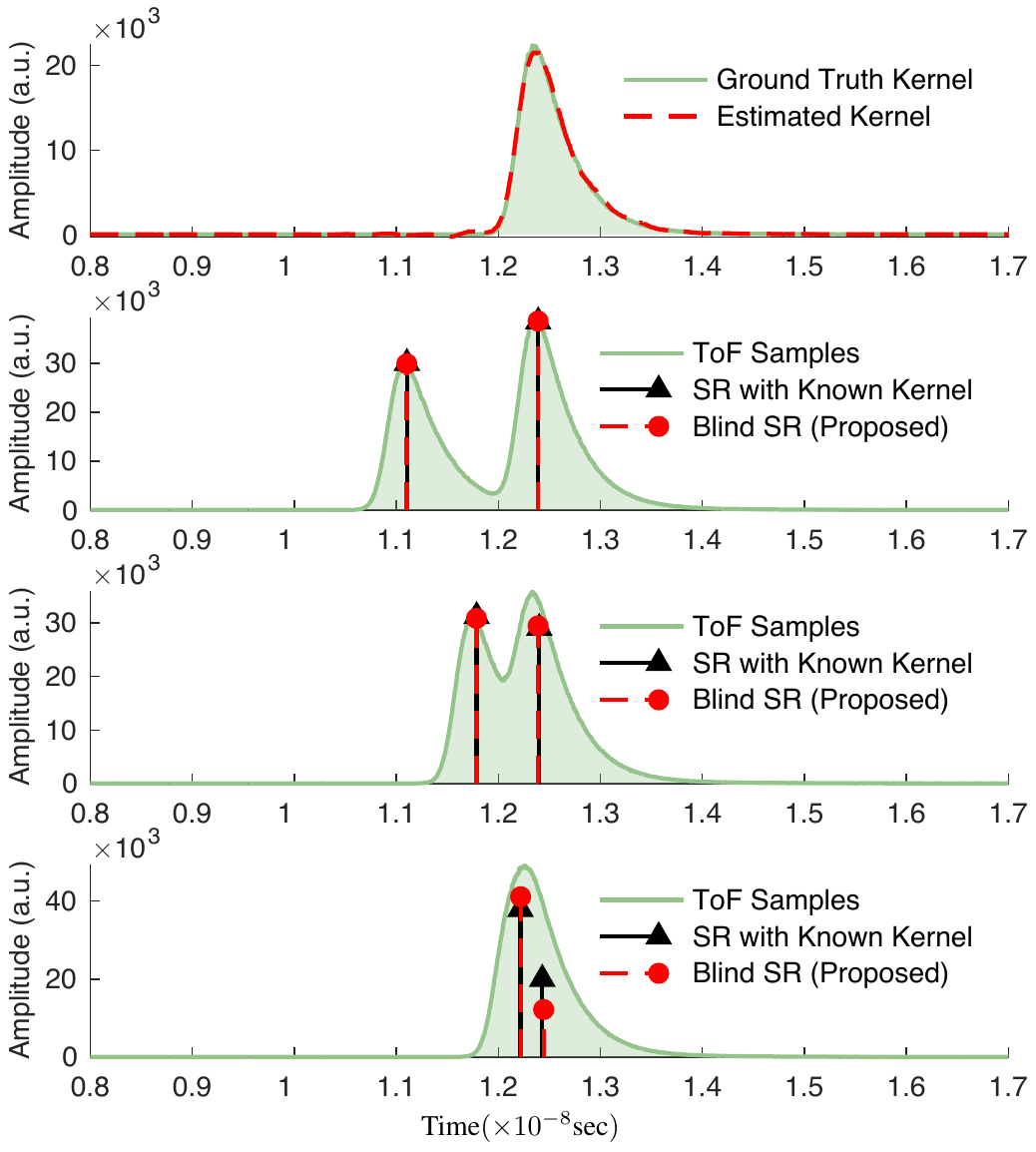}	
\end{center}
\caption{LiDAR based Super-resolved ToF Imaging. The scene consists of two retro-reflecting cubes at distance of $330$ m from the TCSPC system. We show the blind recovery of the kernel and object depths at different inter-reflector separations. Our blind recovery method resolves two reflectors up to a resolution of $13.47$ ps (equivalently $0.20$ cm), showcasing the \SR capability of our method.  
}
\label{fig:spad}
\end{figure}

\begin{figure}[!t]
\begin{center}
\begin{overpic}[width=0.55\columnwidth]{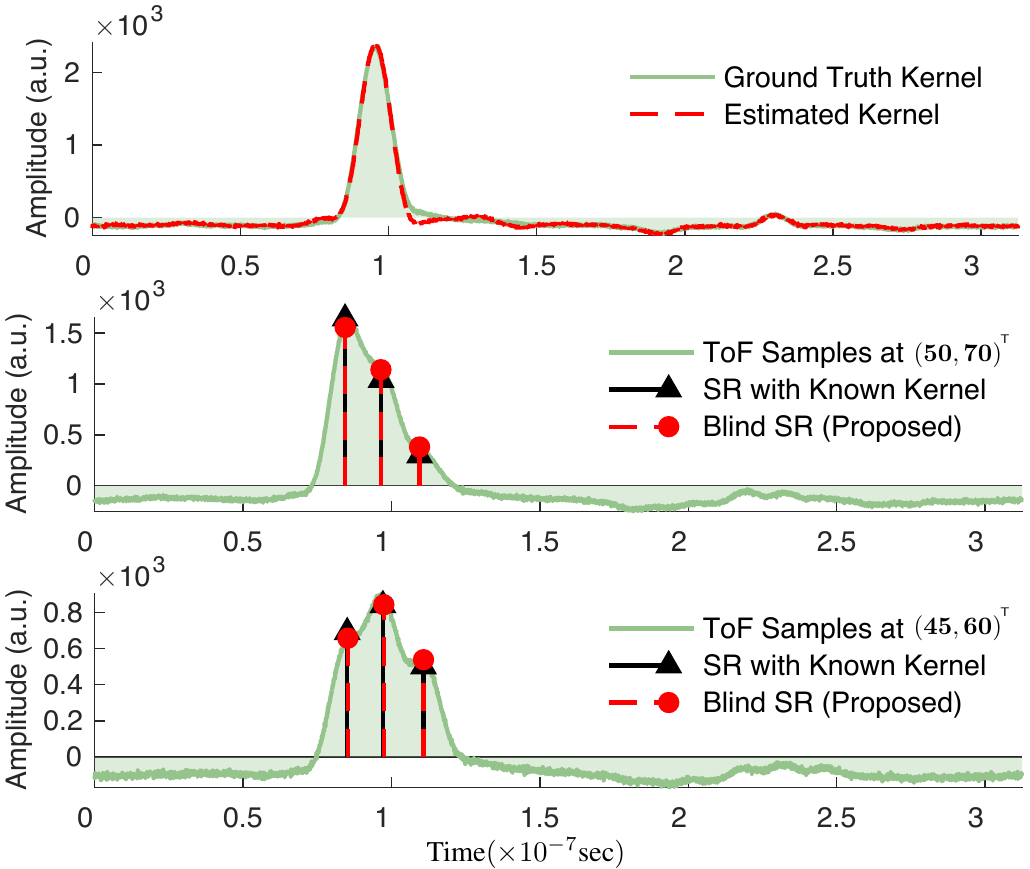}
\put (87.8,49.4) {\colorbox{white}{\sf{\bfseries\fontsize{5pt}{5pt}\selectfont {\color{black} $\left[50\;70\right]^{\top}$ }}}}
\put (87.8,22.6) {\colorbox{white}{\sf{\bfseries\fontsize{5pt}{5pt}\selectfont {\color{black} $\left[45\;60\right]^{\top}$ }}}}
\end{overpic}
\end{center}
\caption{Benchmarking blind ToF imaging for high-order reflections ($K=3$). Two translucent surfaces are placed in front of a wall with separation of $1.8$ and $2$ meters, respectively. The kernel and \srf recovery of two pixel measurements are plotted, showcasing the \SR capability of our method.
}
\label{fig:3spikes}
\end{figure}

\begin{figure}[!t]
\centering
\includegraphics[width=1\columnwidth]{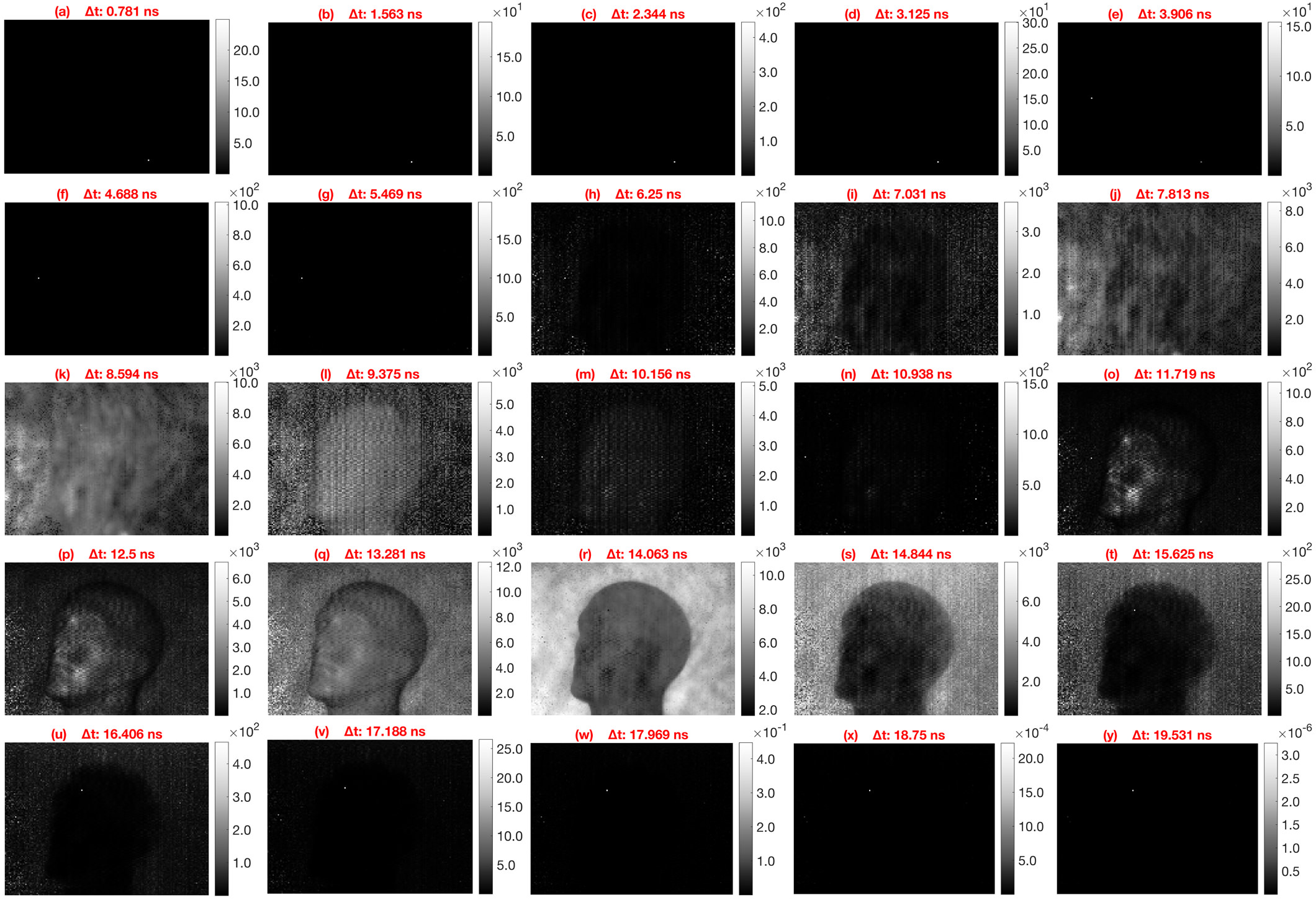} 	
\caption{{Light-in-flight imaging}: The scene consists of a mannequin head placed between a diffusive semi-translucent surface and a wall in the backdrop with a separation of $160$ cm, as shown in \fig{fig:LIF}. 
Our method enables the visualization of the light sweep process without any kernel calibration. 
Light first impinges the diffusive surface in (k), then sweeps over the mannequin head from (o) to (t), and finally reaches the back wall as shown in (y).
We refer to \tabref{tab:LIF} for visualization of light sweep movies corresponding to inter-object separation of $160$, $100$ to $70$ cm, respectively. These movies provide an visual interpretation of pulse propagation through the scene set up in \fig{fig:LIF}.
}
\label{fig:headLIF}
\end{figure}

\subsubsection{Super-resolution}
\label{subsubsec:K=2}

In the context of multi-path imaging, a natural and challenging scenario is to resolve closely-located objects from time profiles. In order to validate the \SR capability of the proposed approach, we perform experiments on two different setups using lock-in sensors and TCSPC-TOF (Time-Correlated Single Photon Counting) systems.  

\bpara{Lock-in Sensors.}
The \ttd scene comprises of a mannequin head placed between a diffusive semi-translucent surface and a wall in the backdrop, as shown in \fig{fig:LIF}. By shifting the translucent surface, the inter-object separation between the diffusive surface and mannequin head is gradually reduced from $160$, $130$, $100$ to $70$ cm, respectively, with $30$ cm reduction in each experiment. This equivalently leads to equidistant spike shifts as shown in \fig{fig:movingSR} (dashed lines). For this dataset, raw data comprising of $120\times 120\times 3968\times 4$ image tensor is acquired from a lock-in ToF sensor, for which $N=3968$ refers to the number of ToF measurements recorded with sampling time $T=96.15$ ps. 

The blind \srf recoveries of a single-pixel ToF measurements ($\mathbf{r} = \left[60\;60\right]^{\top}, K=2$) are presented in \fig{fig:movingSR}. The translucent surface (front) is moving closer to the mannequin head (back), leading to the challenges of separating two objects. This can be intuitively observed from \fig{fig:movingSR} (c) and (d), where one spike is invisible as the inter-object distance is smaller than the kernel width. 
We validate the reconstructed \ttd scene information by comparing the estimated inter-object separation to the actual scene setups. The inter-object separation is computed by $\Delta \widetilde  d_{mk} = | \Delta \widetilde \tau _{m} -\Delta \widetilde \tau _{k} | c/2$, where $c=3\times 10^{8}$\SI{}{\meter/\second} is the speed of light. Here, the $\Delta d_{mk}$ refers to the results using the calibrated kernel. We outline the estimated results as follows:
\begin{enumerate*}[leftmargin = *,itemsep=2pt]
\item $160$ cm in \fig{fig:movingSR} (a): $\Delta \widetilde  d_{01} = 161$ and $\Delta d_{01} = 159$ cm
\item $130$ cm in \fig{fig:movingSR} (b): $\Delta \widetilde  d_{01} = 134$ and $\Delta d_{01} = 131$ cm
\item $100$ cm in \fig{fig:movingSR} (c): $\Delta \widetilde  d_{01} = 98$ and $\Delta d_{01} = 100$ cm
\item $70$ cm in \fig{fig:movingSR} (d): $\Delta \widetilde  d_{01} = 72$ and $\Delta d_{01} = 69$ cm
\end{enumerate*}
This gives rise to the estimated target shift step with $27, 36, 26$ cm (blind recovery) and $28, 31, 31$ cm (known kernel). Compared to $30$ cm shift difference from the actual scene setup, this demonstrates the \SR capability of the proposed blind recovery method. 
The \ttd visualization of the object depth imaging is presented in \fig{fig:headSR}.
We further reconstruct the \ttd scene from the estimated time profiles by running the algorithm for all pixel measurements, as shown in \fig{fig:ToFSR}. 
Despite the challenging experimental setup, the proposed method super-resolves two objects up to a separation uncertainty of $5$ cm, which is equivalent to resolving a separation of $0.33$ ns in time domain. This effectively demonstrates the \SR capability of the proposed method.

\bpara{TCSPC.} 
The experimental setup is borrowed from \cite{HernandezMarin:2007:J,Bhandari:2016:C} and comprises two retro-reflecting corner cubes. In this setup, with one cube fixed, the inter-spacing with the second cube is reduced to create a scenario that requires algorithmic \SR \cite{Bhandari:2016:C}.
For a corner cube, all beams, independent of incident angle, are reflected back in the original direction so the behavior is that of a perfect reflecting surface. In these experiments, the sampling step of the receiver was $T=6.1$ ps, and the collection time for each histogram was $30$ seconds. We plot the estimated time-localized kernel and the raw measurements from three different inter-reflector separations in \fig{fig:spad} (a)-(d), respectively. 
Similar to the scenarios in \fig{fig:movingSR}, the task becomes more challenging as inter-reflector separation gets smaller. 
As shown in \fig{fig:spad}, the reconstructed results are almost indistinguishable between the ground truth (obtained using \cite{Bhandari:2016:C}) and its recovery (see \tabref{tab:exp}), which also matches the experimental setup and reported results \cite{Bhandari:2016:C}. 
It is noteworthy that, the results in \fig{fig:spad} demonstrate the \SR of our method that resolves a separation of $13.47$ ps in time domain (equivalently $0.20$ cm).

\subsubsection{High-Order Imaging}
\label{subsubsec:K=3}

This experiment is dedicated to pushing the limit of our method in the scenarios of high-order multi-path imaging ($K=3$).  
We use a calibrated scene with two translucent surfaces with a wall in the backdrop. The inter-object separation is $1.8$ and $2$ meters, respectively, with a sampling step of $T=70$ ps. 

The blind recovery of two pixel measurements with coordinates $\mathbf{r} = \left[50\;70\right]^{\top}$ and $\mathbf{r} = \left[45\;60\right]^{\top}$ are plotted in \fig{fig:3spikes}. 
It is worth noting that, compared to the data in \fig{fig:ToFkernel} and \fig{fig:movingSR}, the raw ToF measurements in \fig{fig:3spikes} are contaminated by higher levels of noise, creating algorithmic challenges for scene reconstruction. From the experimental results in \tabref{tab:exp}, the estimated inter-object separation from blind recovery is given by:
\begin{enumerate}[itemsep=2pt]
\item Pixel at $\mathbf{r} = \left[45\;60\right]^{\top}$, $\Delta \widetilde  d_{01} = 1.81, \Delta \widetilde  d_{12} = 1.95$ meters;
\item  Pixel at $\mathbf{r} = \left[50\;70\right]^{\top}$, $\Delta \widetilde  d_{01} = 1.82, \Delta\widetilde  d_{12} = 2.01$ meters.
\end{enumerate}
This accurately matches the experimental setup (\ie, $1.8$ and $2$ meters) and the reported results in \cite{Bhandari:2020:J} that used the known kernel, corroborating the effectiveness of the proposed method.

\begin{table}[!t]
\centering
\caption{YouTube Web Links for Light Sweep Movies}
\label{tab:LIF}
\resizebox{0.55\textwidth}{!}{%
\begin{tabular}{@{}ll@{}}
\toprule
Inter-Object Separation & Web Link (YouTube) \\ \midrule
160 cm & \href{https://youtu.be/wMlWJv7B66o}{\texttt{https://youtu.be/wMlWJv7B66o}} \\
100 cm & \href{https://youtu.be/F-g6X85DWO4}{\texttt{https://youtu.be/F-g6X85DWO4}} \\
70 cm &  \href{https://youtu.be/fO_4ivWC2Hg}{\texttt{https://youtu.be/fO\_4ivWC2Hg}}\\ 
\bottomrule
\end{tabular}%
}
\end{table}

\subsection{Light-in-Flight Imaging}
\label{subsec:Light-in-Flight Imaging}

In this experiment, the scene consists of a mannequin head located between a diffusive semi-translucent surface and a wall in the backdrop. The inter-object distance between the mannequin head and the wall is $160$ cm, as demonstrated in \fig{fig:LIF} and \fig{fig:headSR}.  
For this dataset, raw data comprising of $120\times 120\times 3968$ image tensor is acquired from a lock-in ToF sensor with sampling time $T=96.15$ ps. 
Using our custom ToF camera and blind recovery approach, we are able to visualize light sweep over the scene with multi-path effects, as shown in \fig{fig:headLIF}. In the early time-slices, the light first hits the diffusive sheet in \fig{fig:headLIF} (k) ($8.59$ ns). The light then sweeps over the mannequin head on the scene from (o) ($11.72$ ns) to (t) ($15.63$ ns), and eventually hits the back wall in (y) ($19.53$ ns). The time slices correspond to the true geometry of the scene. The light sweep movies for inter-object separation of $160$, $100$ to $70$ cm, respectively, can be visualized via the YouTube links provided in \tabref{tab:LIF}.

\section{Conclusion}
\label{sec: conclusion}

This paper tackles a fundamental algorithmic challenge in optical time-of-flight (ToF) imaging. ToF imaging systems, which are active devices, illuminate scenes with light pulses or kernels and reconstruct them by capturing echoes of back-scattered light. This process enables the understanding of the three-dimensional environment but introduces the challenge of sparse super-resolution, as the echoes exist at a finer time resolution than what conventional digital devices can measure. Additionally, the quality of super-resolution depends on the knowledge of the probing light pulses.

Departing from previous methods, this paper presents an algorithmic strategy that achieves sparse super-resolution without needing information about the emitted light pulses. Our algorithm effectively recovers sparse echoes modeled as continuous-time Dirac impulses, providing a more accurate representation of physical reality. Consequently, our approach eliminates the need for pulse calibration, offering a highly adaptable framework for Blind ToF imaging that extends the limits of time resolution in 3D scene reconstruction. The validation of our approach through extensive hardware experiments, encompassing a variety of scenarios and ToF modalities, demonstrates the empirical robustness and practical benefits of our method. Looking ahead, our algorithmic framework opens several avenues for future research, particularly in expanding this blind approach to additional contexts.

\begin{enumerate}[leftmargin = 25pt]
\item {\bf Algorithmic Frameworks.} Currently, we do not exploit the fact that all pixels utilize the same pulse. This scenario has been explored under the theme of multi-channel sparse blind deconvolution \cite{Kazemi:2014:J,Wang:2016:J} and applying similar priors to our case is highly pertinent. Moreover, considering the blind recovery scenario where each spike is filtered through a distorted version of the original pulse \cite{Bhandari:2017:C} can pave the way for new application areas.

\item {\bf New Sensing Pipelines.} Exploring non-linear sensing modalities leads to the emergence of new classes of inverse problems and provides clear practical advantages over traditional point-wise sampling. In this context, we have identified two potential areas for development: (i) One-bit sensing \cite{Bhandari:2020:J}, which offers a low-complexity implementation, and (ii) the Unlimited Sensing Framework \cite{Bhandari:2020:Ja,Bhandari:2021:J,Bhandari:2022:J}, which simultaneously provides high dynamic range and high digital resolution.

\item {\bf Application Areas.} Our methodology is applicable to a range of challenges including fluorescence lifetime imaging \cite{Bhandari:2015:J}, looking-around-the-corners \cite{Velten:2012:J}, and imaging through scattering \cite{Wu:2012:C}, all of which are significant for practical implementations. Moreover, beyond the realm of optical ToF methods, our approach has potential advantages for related ToF systems such as seismic imaging \cite{Kazemi:2014:J} and radar technologies \cite{DeFigueiredo:1982:J}.

\end{enumerate}

\bibliographystyle{siamplain}

\begin{thebibliography}{10}

\bibitem{Abramson:1978:J}
{\sc N.~Abramson}, {\em Light-in-flight recording by holography}, Opt Lett, 3
  (1978), p.~121.

\bibitem{Adam:2017:J}
{\sc A.~Adam, C.~Dann, O.~Yair, S.~Mazor, and S.~Nowozin}, {\em Bayesian
  time-of-flight for realtime shape, illumination and albedo}, {IEEE} Trans.
  Pattern Anal. Mach. Intell., 39 (2017), pp.~851--864.

\bibitem{Bell:1995:J}
{\sc A.~J. Bell and T.~J. Sejnowski}, {\em An information-maximization approach
  to blind separation and blind deconvolution}, Neural Comput., 7 (1995),
  pp.~1129--1159.

\bibitem{Bhandari:2022:J}
{\sc A.~Bhandari}, {\em Back in the {US}-{SR}: {Unlimited} sampling and sparse
  super-resolution with its hardware validation}, {IEEE} Signal Process. Lett.,
  29 (2022), pp.~1047--1051.

\bibitem{Bhandari:2015:J}
{\sc A.~Bhandari, C.~Barsi, and R.~Raskar}, {\em Blind and reference-free
  fluorescence lifetime estimation via consumer time-of-flight sensors},
  Optica, 2 (2015), p.~965.

\bibitem{Bhandari:2017:C}
{\sc A.~Bhandari and T.~Blu}, {\em {FRI} sampling and time-varying pulses: Some
  theory and four short stories}, in {IEEE} Intl. Conf. on Acoustics, Speech
  and Signal Processing (ICASSP), Mar. 2017.

\bibitem{Bhandari:2017:Ca}
{\sc A.~Bhandari, A.~Bourquard, and R.~Raskar}, {\em Sampling without time:
  Recovering echoes of light via temporal phase retrieval}, in {IEEE} Intl.
  Conf. on Acoustics, Speech and Signal Processing (ICASSP), Mar. 2017.

\bibitem{Bhandari:2020:J}
{\sc A.~Bhandari, M.~H. Conde, and O.~Loffeld}, {\em One-bit time-resolved
  imaging}, {IEEE} Trans. Pattern Anal. Mach. Intell., 42 (2020),
  pp.~1630--1641.

\bibitem{Bhandari:2014:Cb}
{\sc A.~Bhandari, A.~Kadambi, and R.~Raskar}, {\em Sparse linear operator
  identification without sparse regularization? {A}pplications to mixed pixel
  problem in time-of-flight range imaging}, in {IEEE} Intl. Conf. on Acoustics,
  Speech and Signal Processing (ICASSP), May 2014,
  \url{http://dx.doi.org/10.1109/ICASSP.2014.6853619}.

\bibitem{Bhandari:2022:B}
{\sc A.~Bhandari, A.~Kadambi, and R.~Raskar}, {\em Computational Imaging}, MIT
  Press, 1st~ed., Oct. 2022.
\newblock \newline Open Access URL: https://imagingtext.github.io/.

\bibitem{Bhandari:2021:J}
{\sc A.~Bhandari, F.~Krahmer, and T.~Poskitt}, {\em Unlimited sampling from
  theory to practice: {Fourier}-{Prony} recovery and prototype {ADC}}, {IEEE}
  Trans. Sig. Proc.,  (2021), pp.~1131--1141.

\bibitem{Bhandari:2020:Ja}
{\sc A.~Bhandari, F.~Krahmer, and R.~Raskar}, {\em On unlimited sampling and
  reconstruction}, {IEEE} Trans. Sig. Proc., 69 (2020), pp.~3827--3839.

\bibitem{Bhandari:2016:J}
{\sc A.~Bhandari and R.~Raskar}, {\em Signal processing for time-of-flight
  imaging sensors: An introduction to inverse problems in computational 3-d
  imaging}, {IEEE} Signal Process. Mag., 33 (2016), pp.~45--58.

\bibitem{Bhandari:2016:C}
{\sc A.~Bhandari, A.~M. Wallace, and R.~Raskar}, {\em Super-resolved
  time-of-flight sensing via {FRI} sampling theory}, in {IEEE} Intl. Conf. on
  Acoustics, Speech and Signal Processing (ICASSP), Mar. 2016.

\bibitem{Blu:2002:C}
{\sc T.~Blu, H.~Bay, and M.~Unser}, {\em A new high-resolution processing
  method for the deconvolution of optical coherence tomography signals}, in
  {IEEE} Intl. Symp. on Biomedical Imaging (ISBI), {IEEE}, 2002.

\bibitem{Bouman:2022:B}
{\sc C.~A. Bouman}, {\em Foundations of Computational Imaging: {A} Model-Based
  Approach}, Society for Industrial and Applied Mathematics, Jan. 2022.

\bibitem{Candes:2013:J}
{\sc E.~J. Candès and C.~Fernandez‐Granda}, {\em Towards a mathematical
  theory of super‐resolution}, Comm. Pure Appl. Math., 67 (2013),
  pp.~906--956.

\bibitem{Castorena:2015:J}
{\sc J.~Castorena and C.~D. Creusere}, {\em Sampling of time-resolved
  full-waveform {LIDAR} signals at sub-nyquist rates}, {IEEE} Trans. Geosci.
  Remote Sens., 53 (2015), pp.~3791--3802.

\bibitem{Catala:2019:J}
{\sc P.~Catala, V.~Duval, and G.~Peyr\'{e}}, {\em A low-rank approach to
  off-the-grid sparse superresolution}, SIAM Journal on Imaging Sciences, 12
  (2019), pp.~1464--1500.

\bibitem{Claerbout:1973:J}
{\sc J.~F. Claerbout and F.~Muir}, {\em Robust modeling with erratic data},
  Geophysics, 38 (1973), pp.~826--844.

\bibitem{Conde:2020:J}
{\sc M.~H. Conde}, {\em A material-sensing time-of-flight camera}, IEEE Sensors
  Letters, 4 (2020), pp.~1--4.

\bibitem{DeFigueiredo:1982:J}
{\sc R.~J.~P. De~Figueiredo and C.-L. Hu}, {\em Waveform feature extraction
  based on {Tauberian} approximation}, {IEEE} Trans. Pattern Anal. Mach.
  Intell., PAMI-4 (1982), pp.~105--116.

\bibitem{Denoyelle:2019:J}
{\sc Q.~Denoyelle, V.~Duval, G.~Peyr{é}, and E.~Soubies}, {\em The sliding
  {Frank–Wolfe} algorithm and its application to super-resolution
  microscopy}, Inverse Problems, 36 (2019), p.~014001.

\bibitem{Donoho:1992:J}
{\sc D.~L. Donoho}, {\em Superresolution via sparsity constraints}, SIAM
  Journal on Mathematical Analysis, 23 (1992), pp.~1309--1331.

\bibitem{Duarte:2008:J}
{\sc M.~F. Duarte, M.~A. Davenport, D.~Takhar, J.~N. Laska, T.~Sun, K.~F.
  Kelly, and R.~G. Baraniuk}, {\em Single-pixel imaging via compressive
  sampling}, {IEEE} Signal Process. Mag., 25 (2008), pp.~83--91.

\bibitem{Elad:2010:B}
{\sc M.~Elad}, {\em Sparse and redundant representations}, Springer, 2010.

\bibitem{Foix:2011:J}
{\sc S.~Foix, G.~Alenya, and C.~Torras}, {\em Lock-in time-of-flight ({ToF})
  cameras: A survey}, {IEEE} Sensors J., 11 (2011), pp.~1917--1926.

\bibitem{Gao:2014:J}
{\sc L.~Gao, J.~Liang, C.~Li, and L.~V. Wang}, {\em Single-shot compressed
  ultrafast photography at one hundred billion frames per second}, Nature, 516
  (2014), pp.~74--77.

\bibitem{Gariepy:2015:J}
{\sc G.~Gariepy, N.~Krstajić, R.~Henderson, C.~Li, R.~R. Thomson, G.~S.
  Buller, B.~Heshmat, R.~Raskar, J.~Leach, and D.~Faccio}, {\em Single-photon
  sensitive light-in-fight imaging}, Nat. Commun., 6 (2015).

\bibitem{Guo:2022:J}
{\sc R.~Guo, Y.~Li, T.~Blu, and H.~Zhao}, {\em Vector-{FRI} recovery of
  multi-sensor measurements}, {IEEE} Trans. Sig. Proc., 70 (2022),
  pp.~4369--4380.

\bibitem{He:2010:J}
{\sc Z.~He, A.~Cichocki, S.~Xie, and K.~Choi}, {\em Detecting the number of
  clusters in n-way probabilistic clustering}, {IEEE} Trans. Pattern Anal.
  Mach. Intell., 32 (2010), pp.~2006--2021.

\bibitem{Heide:2013:J}
{\sc F.~Heide, M.~B. Hullin, J.~Gregson, and W.~Heidrich}, {\em Low-budget
  transient imaging using photonic mixer devices}, ACM Trans. Graphics, 32
  (2013), pp.~1--10.

\bibitem{HernandezMarin:2007:J}
{\sc S.~Hernandez-Marin, A.~Wallace, and G.~Gibson}, {\em Bayesian analysis of
  lidar signals with multiple returns}, {IEEE} Trans. Pattern Anal. Mach.
  Intell., 29 (2007), pp.~2170--2180.

\bibitem{Jarabo:2017:B}
{\sc A.~Jarabo, B.~Masia, J.~Marco, and D.~Gutierrez}, {\em Recent advances in
  transient imaging: {A} computer graphics and vision perspective}, vol.~1,
  Elsevier {BV}, mar 2017.

\bibitem{Kadambi:2013:J}
{\sc A.~Kadambi, R.~Whyte, A.~Bhandari, L.~Streeter, C.~Barsi, A.~Dorrington,
  and R.~Raskar}, {\em Coded time of flight cameras: sparse deconvolution to
  address multipath interference and recover time profiles}, ACM Trans.
  Graphics, 32 (2013), pp.~1--10.

\bibitem{Kazemi:2014:J}
{\sc N.~Kazemi and M.~D. Sacchi}, {\em Sparse multichannel blind
  deconvolution}, GEOPHYSICS, 79 (2014), pp.~V143--V152.

\bibitem{Kirmani:2014:J}
{\sc A.~Kirmani, D.~Venkatraman, D.~Shin, A.~Colaco, F.~N.~C. Wong, J.~H.
  Shapiro, and V.~K. Goyal}, {\em First-photon imaging}, Science, 343 (2014),
  pp.~58--61.

\bibitem{Kuo:2020:J}
{\sc H.-W. Kuo, Y.~Zhang, Y.~Lau, and J.~Wright}, {\em Geometry and symmetry in
  short-and-sparse deconvolution}, SIAM J. Math. Data Sci., 2 (2020),
  pp.~216--245.

\bibitem{Li:2019:J}
{\sc X.~Li, S.~Ling, T.~Strohmer, and K.~Wei}, {\em Rapid, robust, and reliable
  blind deconvolution via nonconvex optimization}, Appl. Comput. Harmon. Anal.,
  47 (2019), pp.~893--934.

\bibitem{Pellegrini:2000:J}
{\sc S.~Pellegrini, G.~S. Buller, J.~M. Smith, A.~M. Wallace, and S.~Cova},
  {\em Laser-based distance measurement using picosecond resolution
  time-correlated single-photon counting}, Meas Sci Technol, 11 (2000),
  pp.~712--716.

\bibitem{Perrone:2016:J}
{\sc D.~Perrone and P.~Favaro}, {\em A clearer picture of total variation blind
  deconvolution}, {IEEE} Trans. Pattern Anal. Mach. Intell., 38 (2016),
  pp.~1041--1055.

\bibitem{Poon:2019:J}
{\sc C.~Poon and G.~Peyr\'{e}}, {\em Multidimensional sparse super-resolution},
  SIAM Journal on Mathematical Analysis, 51 (2019), pp.~1--44.

\bibitem{Qiao:2015:J}
{\sc H.~Qiao, J.~Lin, Y.~Liu, M.~B. Hullin, and Q.~Dai}, {\em Resolving
  transient time profile in {ToF} imaging via log-sum sparse regularization},
  Optics Letters, 40 (2015), p.~918.

\bibitem{RedoSanchez:2016:J}
{\sc A.~Redo-Sanchez, B.~Heshmat, A.~Aghasi, S.~Naqvi, M.~Zhang, J.~Romberg,
  and R.~Raskar}, {\em Terahertz time-gated spectral imaging for content
  extraction through layered structures}, Nature Communications, 7 (2016).

\bibitem{Repetti:2015:J}
{\sc A.~Repetti, M.~Q. Pham, L.~Duval, E.~Chouzenoux, and J.-C. Pesquet}, {\em
  Euclid in a taxicab: Sparse blind deconvolution with smoothed ${\ell
  _1}/{\ell _2}$ regularization}, {IEEE} Signal Process. Lett., 22 (2015),
  pp.~539--543.

\bibitem{Satat:2015:J}
{\sc G.~Satat, B.~Heshmat, C.~Barsi, D.~Raviv, O.~Chen, M.~G. Bawendi, and
  R.~Raskar}, {\em Locating and classifying fluorescent tags behind turbid
  layers using time-resolved inversion}, Nature Communications, 6 (2015).

\bibitem{Seelamantula:2014:J}
{\sc C.~S. Seelamantula and S.~Mulleti}, {\em Super-resolution reconstruction
  in frequency-domain optical-coherence tomography using the
  finite-rate-of-innovation principle}, {IEEE} Trans. Sig. Proc., 62 (2014),
  pp.~5020--5029.

\bibitem{Sen:2005:C}
{\sc P.~Sen, B.~Chen, G.~Garg, S.~R. Marschner, M.~Horowitz, M.~Levoy, and
  H.~P.~A. Lensch}, {\em Dual photography}, in ACM {SIGGRAPH}, SIGGRAPH05, ACM,
  July 2005.

\bibitem{Shin:2016:J}
{\sc D.~Shin, F.~Xu, F.~N.~C. Wong, J.~H. Shapiro, and V.~K. Goyal}, {\em
  Computational multi-depth single-photon imaging}, Optics Express, 24 (2016),
  p.~1873.

\bibitem{Strang:2011:B}
{\sc G.~Strang and G.~Fix}, {\em A Fourier Analysis of the Finite Element
  Variational Method}, Springer Berlin Heidelberg, 2011.

\bibitem{EHTC:2019:J}
{\sc {The Event Horizon Telescope Collaboration}}, {\em First {M87 Event
  Horizon Telescope} results. iv. {Imaging} the central supermassive black
  hole}, The Astrophysical Journal Letters, 875 (2019).

\bibitem{Toole:2017:C}
{\sc M.~O. Toole, F.~Heide, D.~B. Lindell, K.~Zang, S.~Diamond, and
  G.~Wetzstein}, {\em Reconstructing transient images from single-photon
  sensors}, in {IEEE} Intl. Conf. on Computer Vision and Pattern Recognition
  (CVPR), {IEEE}, jul 2017.

\bibitem{Tur:2011:J}
{\sc R.~Tur, Y.~C. Eldar, and Z.~Friedman}, {\em Innovation rate sampling of
  pulse streams with application to ultrasound imaging}, {IEEE} Trans. Sig.
  Proc., 59 (2011), pp.~1827--1842.

\bibitem{Urigen:2013:J}
{\sc J.~A. Urigen, T.~Blu, and P.~L. Dragotti}, {\em {FRI} sampling with
  arbitrary kernels}, {IEEE} Trans. Sig. Proc., 61 (2013), pp.~5310--5323.

\bibitem{Velten:2012:J}
{\sc A.~Velten, T.~Willwacher, O.~Gupta, A.~Veeraraghavan, M.~G. Bawendi, and
  R.~Raskar}, {\em Recovering three-dimensional shape around a corner using
  ultrafast time-of-flight imaging}, Nature Communications, 3 (2012).

\bibitem{Velten:2013:J}
{\sc A.~Velten, D.~Wu, A.~Jarabo, B.~Masia, C.~Barsi, C.~Joshi, E.~Lawson,
  M.~Bawendi, D.~Gutierrez, and R.~Raskar}, {\em Femto-photography: capturing
  and visualizing the propagation of light}, ACM Trans. Graphics, 32 (2013),
  pp.~1--8.

\bibitem{Velten:2016:J}
{\sc A.~Velten, D.~Wu, B.~Masia, A.~Jarabo, C.~Barsi, C.~Joshi, E.~Lawson,
  M.~Bawendi, D.~Gutierrez, and R.~Raskar}, {\em Imaging the propagation of
  light through scenes at picosecond resolution}, Commun ACM, 59 (2016),
  pp.~79--86.

\bibitem{Wang:2016:J}
{\sc L.~Wang, Q.~Zhao, J.~Gao, Z.~Xu, M.~Fehler, and X.~Jiang}, {\em Seismic
  sparse-spike deconvolution via toeplitz-sparse matrix factorization},
  GEOPHYSICS, 81 (2016), pp.~V169--V182.

\bibitem{Wang:2022:J}
{\sc W.~Wang, J.~Li, and H.~Ji}, {\em $l_1$-norm regularization for
  short-and-sparse blind deconvolution: Point source separability and region
  selection}, SIAM J. Imaging Sci., 15 (2022), pp.~1345--1372.

\bibitem{Wu:2021:J}
{\sc C.~Wu, J.~Liu, X.~Huang, Z.-P. Li, C.~Yu, J.-T. Ye, J.~Zhang, Q.~Zhang,
  X.~Dou, V.~K. Goyal, F.~Xu, and J.-W. Pan}, {\em Non-line-of-sight imaging
  over 1.43 km}, Proceedings of the National Academy of Sciences, 118 (2021).

\bibitem{Wu:2012:C}
{\sc D.~Wu, M.~O{\textquotesingle}Toole, A.~Velten, A.~Agrawal, and R.~Raskar},
  {\em Decomposing global light transport using time of flight imaging}, in
  {IEEE} Intl. Conf. on Computer Vision and Pattern Recognition ({CVPR}),
  {IEEE}, jun 2012.

\end{thebibliography}

\end{document}